\documentclass[12pt]{article}
\usepackage{graphicx}
\usepackage{amsmath}
\usepackage{amssymb}
\usepackage{ulem}
\usepackage{enumerate}
\usepackage{cite}
\usepackage{latexsym}
\begin{document}
\vskip 2cm

\def\warning#1{\begin{center}
\framebox{\parbox{0.8\columnwidth}{\large\bf #1}}
\end{center}}

\begin{center}
{\large {\bf A Massive Field-Theoretic Model for  Hodge Theory}}

\vskip 1.6 cm

{\sf{ \bf S. Krishna$^{(a),}$\footnote{Present Address: Department of Physics, Zakir Husain Delhi College, University of Delhi, \\
$~~~~~~~~~~~~~~~~~~~~~~~~~~~~~$ New Delhi--110002, India.}, 
R. Kumar$^{(b),}$\footnote{Corresponding author.}, R. P. Malik$^{(c,d)}$}}\\
\vskip .1cm
{\it $^{(a)}$ Indian Institute of Science Education and Research Mohali,\\ Sector 81, SAS Nagar, Manauli, Punjab--140306, India}\\
\vskip .1cm
{\it $^{(b)}$Department of Physics \& Astrophysics,\\ University of Delhi, New Delhi--110007, India}\\
\vskip .1cm
{\it $^{(c)}$Physics Department, CAS, Institute of Science,\\
Banaras Hindu University (BHU), Varanasi--221005,  India}\\
\vskip .1cm
{\it $^{(d)}$DST--CIMS, Institute of Science, BHU, Varanasi--221005, India}\\
\vskip .2cm
{E-mails: skrishna.bhu@gmail.com; raviphynuc@gmail.com;  rpmalik1995@gmail.com }\\
\end{center}

\vskip 1.4 cm

\noindent
{\bf Abstract:} Within the framework of   Becchi--Rouet--Stora--Tyutin (BRST) formalism, we show that the four 
$(3+1)$-dimensional (4D) {\it  massive} Abelian 2-form gauge theory (without any interaction with matter fields) 
is a model for the Hodge theory because its {\it discrete} and {\it continuous} symmetry transformations (and their 
corresponding Noether conserved charges) provide the physical realizations of the de Rham cohomological 
operators of differential geometry at the {\it algebraic} level. For this purpose, we incorporate  
the {\it pseudo-scalar} and {\it axial-vector} fields which appear in the theory with {\it negative } kinetic terms 
({\it but} with proper definition of {\it mass}). The negative kinetic terms, for the above fields, are {\it essential} 
so that our theory could respect the {\it discrete} symmetry transformations which provide the physical 
realizations of the Hodge duality operation in the domain of differential geometry. Thus, our present endeavour, not {\it only} provides the physical 
realizations of {\it all} the mathematical ingredients connected with the de Rham cohomological operators of differential geometry, 
it also sheds  light on the existence and emergence of fields with negative kinetic terms. We discuss the implications and relevance 
of the latter fields in the context of current models of dark matter and dark energy as well as the bouncing models of Universe.

\vskip 1cm
\noindent
PACS numbers: 11.15.-q;  11.30.-j;  03.70.+k; 95.35.+d

\vskip 0.5cm
\noindent
{\it Keywords:} 4D massive Abelian 2-form gauge theory; Nilpotent symmetries; Bosonic symmetry; Massive model of Hodge theory; 
Fields with negative kinetic term; Cosmological models of Universe


\newpage


\section{Introduction} \label{Sect. 1}

The standard model of particle physics, based on the fundamental principles of {\it local} 
gauge invariant  (non-)Abelian 1-{\it form} gauge theories,
is one of the most successful theories of high energy physics where  there is a stunning degree of agreement between theory and experiment.  
This model also provides the theoretical framework for the unification of electromagnetic, weak and strong interactions of nature. 
However, one has to go beyond the purview of standard model of particle physics in view of the fact that the neutrinos have been 
found to be {\it massive} by precise experimental observations. This experimental result is one of the many crucial reasons that 
has compelled theoretical physicists to propose $new$ models in high energy physics that are {\it mostly} based on the ideas of 
supersymmetry (e.g. supersymmetric models of quantum field theories and superstring theories). One of the hottest candidates, 
in this direction, is the basic ideas behind (super)string theories which lead to the theoretical 
description of quantum gravity.  These theories {\it also} provide a  
theoretical framework for the unification of {\it all} four fundamental forces of nature. In the quantum excitations of the superstring 
theories, the higher $p$-form ($p = 2, 3, 4,...$) gauge fields appear very naturally thereby going beyond the realm of standard model 
of particle physics in a {\it subtle} manner (because the latter theoretical model, as stated earlier, is based {\it only} on the basic principles 
of local gauge invariant (non-)Abelian 1-{\it form} theories). Thus, the study of higher $p$-form ($p = 2, 3, 4,...$) gauge theories has become quite 
interesting and important during the last few years due to its connection with 
the (super)strings and their quantum excitations.

In the covariant canonical quantization of {\it gauge and reparametrization invariant  theories} of any kind, the role of Becchi--Rouet--Stora--Tyutin 
(BRST) formalism \cite{brs1,brs2,brs3,brs4} is quite crucial as it maintains {\it unitarity} and ``quantum" gauge (i.e. BRST) invariance 
at any arbitrary order of perturbative computations for any physically allowed process. We have established, in our earlier 
works (see, e.g. for a brief review \cite{rpm0,rk1}), that any arbitrary Abelian $p$-form ($p = 1, 2, 3,...$) gauge theory,  in  $\text{D}= 2p$ dimensions 
of spacetime, is endowed with the (anti-)BRST as well as (anti-)co-BRST symmetries within the framework of BRST formalism. Such 
theories have been shown to provide a set of  tractable {\it physical} examples for the Hodge theory where the symmetries 
(and corresponding conserved charges) provide the {\it physical} realizations of the de Rham cohomological operators of differential 
geometry \cite{eug,mukh,van,des}. In our earlier works (see, e.g. \cite{rpm0,rk1,rpm1,rpm2,rpm3,rpm4,rpm5,rpm6}), we have established 
that the 2D (non-)Abelian 1-form gauge theories, 4D Abelian 2-form and 6D Abelian 3-form gauge theories provide the examples 
of Hodge theory. Such studies are {\it physically} important because we have shown that the 2D (non-)Abelian 1-form gauge 
theories provide a set of {\it new} models of topological field theories (TFTs) \cite{rpm7} which capture a few aspects of 
Witten-type TFTs \cite{witt} and some salient features of Schwarz-type TFTs \cite{schw}. In addition, it has been shown 
that the {\it free} 4D Abelian 2-form and 6D Abelian 3-form gauge theories are the examples of quasi-TFTs \cite{rpm8,rk1}. 
An {\it interacting} Abelian 1-form gauge theory (with $massless$  Dirac fields) has {\it also} been shown to be a perfect model of
Hodge theory \cite{rpm9} because of its various discrete and continuous symmetries and their connections with 
the {\it algebra} of de Rham cohomological operators of differential geometry (including the Hodge duality operation).

All the theories, that have been mentioned in the previous paragraph, are {\it massless} Abelian $p$-form gauge theories which have 
been shown to be the models for the Hodge theory in $\text{D} = 2p$ dimensions of spacetime (within the framework of BRST formalism 
\cite{rpm0,rk1}). In our earlier work \cite{rpm10}, for the first time, we have demonstrated  that the St\"{u}ckelberg modified 2D Proca 
theory (i.e. a {\it massive} 2D Abelian 1-form gauge theory) is {\it also} a model for the Hodge theory provided we invoke a {\it new} field in the 
theory (which is nothing {\it but} a pseudo-scalar field that turns up in the theory with a {\it negative} kinetic term). The 
continuous and discrete symmetries of the theory enforce the scalar field of the theory to possess the {\it positive} kinetic term but 
the pseudo-scalar field of the theory, as pointed out earlier, is forced to acquire a {\it negative} kinetic term (with a {\it properly} 
well-defined mass). Hence, the latter field mimics  one of the key properties of the dark matter which is quite  popular in modern 
literature \cite{carr,sah,bert,cald}. Thus, the 2D St\"{u}ckelberg modified Proca theory (i.e. a {\it massive} 2D Abelian 1-form gauge theory) 
provides a theoretical basis and motivation to look for the discussion of existence and emergence of fields with negative kinetic terms in the 
{\it physical} four $(3+1)$-dimensional (4D) theories within the framework of quantum field theory (QFT) where the BRST formalism plays a 
crucial role (as far as the symmetry properties and their conserved charges are concerned).

The central theme of our present investigation is to carry forward the ideas  \cite{rpm10} of 2D St\"{u}ckelberg  modified {\it massive} 
Abelian 1-form gauge theory (i.e. the modified Proca theory) to the four $(3+1)$-dimensional {\it massive} Abelian 2-form gauge 
theory and demonstrate the existence of axial-vector and pseudo-scalar fields which turn up with {\it negative} kinetic terms 
(but with well-defined mass as they satisfy the Klein--Gordon equation). In fact, the {\it symmetries} of the St\"{u}ckelberg  modified   
{\it massive} 4D  Abelian 2-form gauge theory are such that they {\it fix} all the signatures of {\it all} the terms that appear in the 
coupled (but equivalent) Lagrangian densities. These symmetries are responsible for the 
proof of  this {\it massive} physical 4D model to become an example of Hodge theory within the framework of BRST formalism. To be precise, 
we have {\it six} continuous symmetries in the theory, out of which, four are fermionic (supersymmetric-type) and {\it two} of 
them are {\it bosonic} in nature. We have shown that the {\it algebra} of continuous symmetry transformation operators (and corresponding 
conserved charges) obey exactly the {\it same} algebra as the {\it algebra} of de Rham cohomological operators of differential geometry. 
In addition to the above {\it six} continuous symmetries, we have also shown the existence of {\it two} appropriate discrete symmetries in the theory which 
provide the physical realizations of the Hodge duality operation of differential geometry at the {\it algebraic} level in the 
well-known relationship between the co-exterior derivative and exterior derivative. As far as the  {\it physical} consequences of 
our present study is concerned, we observe that the emergence of the fields/particles with the negative kinetic terms as one of the possible candidates of dark matter/dark 
energy. This result  is the culmination of {\it all} our earlier works \cite{rpm0,rk1,rpm1,rpm2,rpm3,rpm4,rpm5,rpm6} where we have proposed 
the existence of 4D and 6D quasi-TFTs and a couple of $new$ models for the 2D TFTs within the framework of BRST 
formalism (see, e.g. \cite{rpm0,rk1,rpm1,rpm2,rpm3,rpm4,rpm5,rpm6,rpm7} for details).

Against the backdrop of our discussions in the previous paragraphs, we would like to say a few things about one of the the modern 
theoretical understandings of the possible candidates for the dark matter and dark energy \cite{carr,sah,bert,cald}. The 
pressing problems of theoretical physics of modern times is to explain the accelerated expansion of our Universe which has 
been established by several  experimental observations \cite{spe,ries,perl,teg,aba1,aba2}. The idea of the existence of 
dark energy has been invoked to explain the accelerated expansion (of our present Universe). During the past few years, 
the fields/particles with {\it negative} 
kinetic terms have been considered by many theoretical and experimental researchers as the one of the possible candidates for the dark matter and 
dark energy \cite{cald1,gib,carr1,zhu,feng}. One of the central outcomes of our present investigation is to demonstrate the existence 
of a {\it massive} pseudo-scalar and an axial-vector fields in the discussions of the {\it massive} 4D Abelian 2-form gauge theory where 
the above fields (with {\it negative} kinetic terms)  appear due to the {\it symmetry} considerations {\it alone}.  In fact, in our earlier 
works on 2D Proca theory \cite{rpm10}, we have established the existence of a {\it massive} pseudo-scalar field with negative 
kinetic term (see, also Appendix A) which is required in the proof of {\it this} theory to be a model for the Hodge theory. It 
is but natural to conclude that, in the {\it massless} limit, the above pseudo-scalar field becomes a possible candidate for the dark 
energy. Thus, in our present endeavour, we provide the {\it unified} theoretical explanation for the possible existence and emergence 
of the fields corresponding to the dark matter and dark energy within the framework of QFT where the BRST formalism plays a decisive role. 

The following motivating factors have been at the heart of our present investigation. First and foremost, so far, 
we have been able to prove the 2D Proca (i.e. a {\it massive} Abelian 1-form) theory to be an example of Hodge theory \cite{rpm10}.
Thus, it has been a challenging problem  for us to prove a massive {\it physical} 4D Abelian 2-form theory to be a model for the Hodge theory. We 
have accomplished this goal in our present endeavour. Second, in our earlier work \cite{rpm5}, we have shown that the
4D free Abelian 2-form {\it gauge} theory is a model for the Hodge theory. Thus, it has been a tempting and interesting problem 
for us to prove the {\it massive} version of the above 4D theory to be a model for the Hodge theory, too. We have achieved 
this objective in our present investigation. Finally, the underlying mathematical/theoretical exercises (connected with the 
proof of the  models to be the examples of  Hodge theory) have been done by us for the 1D, 2D, 4D and 6D theories which are 
nothing but the toy models in 1D \cite{rpm11,rk3} as well as the field theoretical systems 
\cite{brs2, rpm3, rpm4, rpm5, rpm6, rpm7,rpm8,rpm9,rpm10} in various other dimensions. It has been a challenge for 
us to show the {\it physical} implications of these studies. In our present investigation, we have demonstrated that such 
studies lead to the emergence of fields/particles with {\it negative } kinetic terms which might be, perhaps, one of the possible candidates for 
the dark matter and dark energy \cite{carr,sah,bert,cald} within the framework of BRST  formalism.

The contents of our present investigation are organized as follows. First of all, we discuss the bare essentials 
of the St{\"u}ckelberg  approach to convert the {\it massive} 4D Abelian 2-form theory (endowed with second-class constraints) 
into a gauge theory (endowed with first-class constraints) by adding {\it some} extra fields (i.e.  the analogue of the usual 
St{\"u}ckelberg's field) in Section \ref{Sect. 2}. The linearized version of the coupled (but equivalent) Lagrangian densities (that respect 
the (anti-)BRST symmetry transformations {\it together}) are discussed in Section \ref{Sect. 3}. Our Section \ref{Sect. 4} deals with the discussions on the off-shell 
nilpotent (anti-)co-BRST symmetry transformations. In  Section \ref{Sect. 5}, we elaborate on the existence of a {\it unique} bosonic symmetry 
transformation for our  (anti-)BRST and (anti-)co-BRST symmetry  invariant Lagrangian densities. In Section \ref{Sect. 6}, we discuss the 
existence of the ghost-scale symmetry and {\it discrete} symmetry transformations. Our Section \ref{Sect. 7} deals with the algebraic structures 
of all the continuous symmetry transformations (and corresponding conserved Noether charges) where we establish their connection 
with the {\it algebra} of the de Rham cohomological operators. In Section \ref{Sect. 8}, we concisely comment on the fields with $negative$ kinetic terms 
which are the possible candidates for the dark matter/dark energy.   Finally,  we make some concluding remarks in Section \ref{Sect. 9} 
and point out a few future directions for further investigation(s). In this section, we also mention the physical implications 
of the fields with negative kinetic term in the context of cosmological models.

In our Appendix A, we briefly mention the ideas behind the existence of a pseudo-scalar field with {\it negative } kinetic term
in the context of a 2D Proca theory (which is a precursor to our discussions on our present 4D massive Abelian 2-form theory). 
Our Appendix B is devoted to the discussion of change in the kinetic term ($\frac{1}{12}\, H^{\mu\nu\eta} H_{\mu\nu\eta}$) 
for the gauge field $B_{\mu\nu}$ due to the redefinition of the gauge field
$B_{\mu\nu}$ (cf. Eq. (\ref{2}) below). In our Appendix C, we demonstrate diagrammatically the existence of the CF-type restrictions for our 
model of a  4D St{\"u}ckelberg modified massive gauge theory .

\paragraph{Convention and notations:} We adopt the convention of the {\it left-derivative} w.r.t. all the {\it fermionic} fields of our
theory in appropriate/relevant computations. The background {\it flat} metric tensor for the 4D Minkowskian spacetime manifold is 
chosen to be: $\eta_{\mu\nu} \equiv \eta^{\mu\nu}$ = diag $(+1, -1, -1, -1)$ so that for a non-null vector $A_\mu$, the dot product 
$\partial \cdot A = \eta^{\mu\nu} \partial_\mu A_\nu= \partial_0 A_0 - \partial_i A_i \equiv \eta_{\mu\nu} \partial^\mu A^\nu$ 
where the Greek indices $\mu, \nu, \lambda,... = 0, 1, 2, 3$ correspond to the time and space directions and the Latin indices 
$i ,j, k,... = 1, 2, 3$ stand for the space directions {\it only}. We choose the 4D Levi-Civita tensor $\varepsilon_{\mu\nu\lambda\kappa}$ 
such that $\varepsilon_{0123} = +1 = - \varepsilon^{0123}$ and $\varepsilon_{\mu\nu\lambda\kappa} \varepsilon^{\mu\nu\lambda\kappa}= - \,4!$,
$\varepsilon_{\mu\nu\lambda\kappa} \varepsilon^{\mu\nu\lambda\rho}= - \,3! \,\delta^\rho_\kappa$, etc., and 
$\varepsilon_{0ijk} = - \varepsilon^{0ijk} \equiv \epsilon_{ijk}$ is the 3D Levi-Civita tensor. We adopt the notations 
$s_{(a)b}$ and  $s_{(a)d}$ for the nilpotent (anti-)BRST and (anti-)co-BRST [i.e (anti-)dual BRST] transformations 
(and corresponding charges are denoted by $Q_{(a)b}$ and $Q_{(a)d}$) in the whole body of our text. These transformations (i.e. $s_{(a)b}$ 
and $s_{(a)d}$) are supersymmetric-type in nature as they transform bosonic fields into fermionic fields and {\it vi\`{c}e-versa}. 
We also choose the convention of derivative w.r.t. the second-rank antisymmetric tensor field as: 
$(\partial B_{\mu\nu}/\partial B_{\rho\sigma}) = \frac{1}{2!}\,\big(\delta^\rho_\mu \,
\delta^\sigma_\nu - \delta^\rho_\nu \, \delta^\sigma_\mu \big)$, etc.

\paragraph{Standard definitions:} We briefly mention here the basic concepts behind the key definitions of a few aspects of 
differential geometry that are needed for the full appreciation of our present work:
\begin{enumerate}
\item {\it de Rham cohomological operators}: On a compact manifold without a boundary, we define a set of {\it three} operators 
($d$, $\delta$, $\Delta$) which are christened as the exterior derivative operator, co-exterior derivative operator and 
Laplacian operator, respectively. These operators   follow an algebra: $d^2 = 0$, $\delta^2 = 0$, $\Delta = \{d, \delta\}$, 
$[\Delta, d] = [\Delta, \delta] = 0$ which is popularly   known as the Hodge algebra where the (co-)exterior derivatives 
$(\delta)d$ are connected by the relationship: $\delta = \pm \, * \,d\,*$. Here $*$ is nothing but the Hodge duality 
operation (on a {\it given} compact manifold without a boundary).
\vskip .3cm
\item {\it Hodge decomposition theorem:} On the manifold discussed above, any arbitrary form $f_n$ (of degree $n$) 
can be {\it uniquely} written as the sum of a harmonic form $(\omega_n)$, an exact form $(e_n)$ and a co-exact form $(c_n)$ as
$$f_n = \omega_n + e_n + c_n,$$
where $e_n = d g_{n-1}$, $c_n = \delta h_{n+1}$. Here $g_{n-1}$ and  $h_{n+1}$ are the non-zero forms of degree $(n-1)$ and $(n+1)$,
respectively. In other words, we have the following
$$f_n = \omega_n + d \;g_{n-1} + \delta\; h_{n+1},$$ 
where $\omega_n$ is the harmonic form (i.e. $\Delta \omega_n = 0 \Rightarrow d \omega_n = 0$ and $\delta \omega_n = 0$).
\end{enumerate}


\section {Preliminaries: Lagrangian formulation} \label{Sect. 2}

We begin with the four $(3+1)$-dimensional (4D) Kalb--Ramond Lagrangian density \cite{ogi,kr,kaul} for the free Abelian 2-form  
{\it  massive} theory (with rest mass $m$) as (see, e.g. \cite{rk4} for details)
\begin{eqnarray}
{\cal L}_{(0)} = \frac{1}{12}\, H^{\mu\nu\eta} H_{\mu\nu\eta} - \frac{m^2}{4}\, B^{\mu\nu} B_{\mu\nu},
\label{1}
\end{eqnarray}
where the antisymmetric $(B_{\mu\nu} = - B_{\nu\mu})$ tensor field $B_{\mu\nu}$ is the 4D Abelian 2-form 
$\big(B^{(2)} = \frac{1}{2!}\,(dx^\mu \wedge dx^\nu) \, B_{\mu \nu} \big)$ gauge field and the curvature (i.e. field strength) tensor 
$H_{\mu\nu\eta} = \partial_\mu B_{\nu\eta} + \partial_\nu B_{\eta\mu} + \partial_\eta B_{\mu\nu}$ is derived 
from the 3-form $ \big(H^{(3)} = d B^{(2)} \equiv \frac{1}{3!}\,(dx^\mu \wedge dx^\nu \wedge dx^\eta)\, H_{\mu\nu\eta} \big)$.
It is clear that the mass dimension of $B_{\mu\nu}$ is $[M]$ in the natural units ($\hbar = c = 1$) for the 4D theory. Because 
of the presence of  mass term $(- \frac{m^2}{4}\, B_{\mu\nu}\, B^{\mu\nu})$, there is {\it no} gauge invariance at this stage 
because the above Lagrangian density is endowed with the second-class constraints (see, e.g. \cite{rk4}) in the terminology of 
Dirac's prescription for the classification scheme \cite{dira,sund}. We note that the Euler-Lagrange equation of motion (EL-EOM) from
${\cal L}_{(0)}$ is: $\partial_\mu H^{\mu\nu\eta} + m^2\, B^{\nu\eta} = 0$. It is clear that we obtain the usual Klein--Gordan 
equation [i.e. $ (\Box + m^2)\, B_{\mu\nu} = 0$] for the massive Abelian 2-form field $(B_{\mu\nu})$ because we note that the 
EL-EOM: $\partial_\mu H^{\mu\nu\eta} + m^2\, B^{\nu\eta} = 0$ implies that $\partial_\mu B^{\mu\nu} = \partial_\nu B^{\mu\nu} = 0$. 
The latter conditions are {\it true} (i.e. $\partial_\nu \partial_\mu H^{\mu\nu\eta} + m^2\, \partial_\nu B^{\nu\eta} = 0 
\Rightarrow \partial_\nu B^{\nu\eta} = 0$)
because, for the {\it massive} Abelian 2-form theory, we note that the rest mass $m \neq 0$.

Using the St{\"u}ckelberg's technique, it can be checked that, we can have the following modification/redefinition  
for the antisymmetric tensor field $B_{\mu\nu}$    
\begin{eqnarray}
B_{\mu\nu} &\to& B_{\mu\nu} - \frac{1}{m}\, \Phi_{\mu\nu} - \frac{1}{2\,m}\, \varepsilon_{\mu\nu\eta\kappa}\, \tilde \Phi^{\eta\kappa} \nonumber\\
&\equiv& B_{\mu\nu} - \frac{1}{m}\, \big(\partial_\mu \phi_\nu - \partial_\nu \phi_\mu + \varepsilon_{\mu\nu\eta\kappa}\, \partial^\eta \tilde \phi^\kappa\big)\nonumber\\
&\equiv& B_{\mu\nu} - \frac{1}{m}\, \Phi_{\mu\nu} - \frac{1}{m}\, {\cal F}_{\mu\nu},
\label{2}
\end{eqnarray}
where the Abelian 2-form $\Phi^{(2)} = \frac{1}{2!}\,(dx^\mu \wedge dx^\nu)\,\Phi_{\mu\nu} \equiv d \,\Phi^{(1)}$ 
(with vector 1-form  $\Phi^{(1)} = dx^\mu\, \phi_\mu$, 
$\Phi_{\mu\nu} = \partial_\mu \phi_\nu - \partial_\nu \phi_\mu$) is constructed from a vector field $\phi_\mu$. On the contrary, the   dual
antisymmetric tensor $\tilde \Phi_{\mu\nu} = \partial_\mu \tilde \phi_\nu - \partial_\nu \tilde \phi_\mu$ is constructed with the help of an
axial-vector $\tilde \phi_\mu$ which is derived from the axial-vector 1-form $\tilde \Phi^{(1)} = dx^\mu \,\tilde \phi_\mu$. 
To make the parity of $B_{\mu\nu}$, $\Phi_{\mu\nu}$ and  $\tilde \Phi_{\mu\nu}$ on equal footing, we have taken, in Eq. (\ref{2}), the following
\begin{eqnarray} 
* \,d\, \tilde \Phi^{(1)} = \frac{1}{2!}\, (dx^\mu \wedge dx^\nu)\, {\cal F}_{\mu\nu}, 
\label{3}
\end{eqnarray}  
where ${\cal F}_{\mu\nu} = \frac{1}{2!}\, \varepsilon_{\mu\nu\eta\kappa}\, \tilde \Phi^{\eta\kappa}$ and $*$ is the Hodge 
duality operation on a 2-form $(d \,\tilde \Phi^{(1)})$ which is defined on a flat 4D Minkowskian spacetime 
manifold\footnote{We would like to mention here that, in our earlier work on the local duality invariance of the source free  Maxwell's 
equations with two potentials \cite{prad}, we have defined the field strength tensor as: $F_{\mu\nu} = \partial_\mu V_\nu 
- \partial_\nu V_\mu + \varepsilon_{\mu\nu\eta\kappa}\, \partial^\eta A^\kappa$ where $V_\mu$ and $A_\mu$ are the vector 
and axial-vector 1-form potentials, respectively.}. It is straightforward to check that the 
Lagrangian density (\ref{1}) transforms to the following \cite{rk4}
\begin{eqnarray}
{\cal L}_{(0)} \to {\cal L}_{(1)} &=& \frac{1}{12}\, H^{\mu\nu\eta} H_{\mu\nu\eta} - \frac{m^2}{4}\, B^{\mu\nu} B_{\mu\nu}
- \frac{1}{4}\, \Phi^{\mu\nu} \Phi_{\mu\nu} \nonumber\\
&+& \frac{1}{4}\, \tilde \Phi^{\mu\nu} \tilde \Phi_{\mu\nu}
+ \frac{m}{2}\, B^{\mu\nu} \Phi_{\mu\nu} 
+ \frac{m}{4}\, \varepsilon^{\mu\nu\eta\kappa} \, B_{\mu\nu} \tilde \Phi_{\eta\kappa}, 
\label{4}
\end{eqnarray}
(modulo some total spacetime derivative terms) under the modification (\ref{2}). It should be noted that the kinetic term 
(i.e. $\frac{1}{12}\, H^{\mu\nu\eta} H_{\mu\nu\eta}$) does {\it not} change\footnote{We discuss, in detail, the key 
mathematical and physical ingredients  about this claim in our Appendix B.} in a meaningful manner under the redefinition/modification 
(\ref{2}). For our 4D theory, it is straightforward to note that the mass dimension of fields $\phi_\mu$ and $\tilde \phi_\mu$ 
is $[M]$ in the natural unit where $\hbar = c =1$.  The above Lagrangian density respects (i.e. $\delta^{(1)}_g {\cal L}_{(1)} = 0$) 
the following ``scalar" gauge transformation $\delta_g^{(1)}$, namely;
\begin{eqnarray}
\delta^{(1)}_g \phi_\mu = \partial_\mu \Sigma, \qquad \delta^{(1)}_g \tilde \phi_\mu = \partial_\mu  \tilde \Sigma, 
\qquad \delta^{(1)}_g B_{\mu\nu} = 0,  
\label{5}
\end{eqnarray}
where $\Sigma$ is a scalar and $\tilde \Sigma$ is a pseudo-scalar {\it local} gauge transformation parameters. 
In addition, it also respects the following other symmetry (i.e. ``tensor" gauge symmetry) transformations
\begin{eqnarray} 
\delta^{(2)}_g \phi_\mu = -\, m\, \Lambda_\mu, \qquad \delta^{(2)}_g \tilde \phi_\mu = 0, 
\qquad \delta^{(2)}_g B_{\mu\nu}  = - \,(\partial_\mu  \Lambda_\nu - \partial_\nu \Lambda_\mu),
\label{6}
\end{eqnarray}   
where $\Lambda_\mu$ is a {\it local} Lorentz vector gauge transformation parameter.
To be more precise, it can be explicitly checked that 
$\delta^{(2)}_g {\cal L}_{(1)} = \partial_\mu\big[- m \, \varepsilon^{\mu\nu\eta\kappa}\, \Lambda_\nu \, 
(\partial_\eta \tilde \phi_\kappa)\big]$. Thus, the action integral $S = \int d^4x\, {\cal L}_{(1)}$ remains 
invariant under $\delta^{(2)}_g$ for physically well-defined fields that vanish-off at $\pm\,\infty$. 
These {\it ``classical"} continuous  symmetry transformations (\ref{5}) and (\ref{6}) would play very important roles in our later discussions 
on the subject of off-shell nilpotent and absolutely anticommuting (anti-)BRST  symmetries which are {\it ``quantum"} in nature (cf. Section \ref{Sect. 3}).

We shall focus on the form of the Lagrangian density ${\cal L}_{(1)}$  for our further discussions within the framework 
of BRST formalism where we shall discuss the (anti-)BRST and (anti-)co-BRST symmetry transformations (cf. Sections \ref{Sect. 3} and \ref{Sect. 4}) which are {\it basic} 
ingredients to prove the present 4D {\it massive} Abelian 2-form gauge theory to be a field-theoretic  model for the Hodge theory.  
We note that the Lagrangian density ${\cal L}_{(1)}$ is {\it singular} w.r.t. all the {\it three} basic fields 
($B_{\mu\nu},\, \phi_\mu, \tilde \phi_\mu$) of the theory (see, e.g. \cite{dira,sund}). Thus, for the BRST 
quantization of the theory, we have to add the gauge-fixing terms which have their origin in the co-exterior derivative 
$\delta = - * d *$ (where $*$ is the Hodge duality operator  on the 4D flat Minkowskian spacetime manifold and minus sign 
has been taken because our background spacetime manifold is an {\it even} dimensional). It can be readily checked 
that we have the following:
\begin{eqnarray}
&& \delta\, B^{(2)} = - * d* B^{(2)} = dx^\mu (\partial^\nu B_{\nu\mu}), \nonumber\\
&& \delta \,\phi^{(1)} = - * d* \phi^{(1)}  = (\partial^\mu \phi_\mu)\equiv \big( \partial \cdot \phi\big), \nonumber\\
&&\delta \,\tilde \phi^{(1)} = - * d* \tilde \phi^{(1)}  = (\partial^\mu \tilde \phi_\mu) \equiv \big(\partial \cdot \tilde \phi \big).
\label{7}
\end{eqnarray}
Thus, the Lagrangian density ${\cal L}_{(1)}$ is modified and generalized to ${\cal L}_{(2)}$ as: 
\begin{eqnarray}
{\cal L}_{(1)} \to {\cal L}_{(2)} &=& {\cal L}_{(1)}  + \frac{1}{2} (\partial^\nu B_{\nu\mu}) (\partial_\rho B^{\rho\mu}) 
- \frac{1}{2} (\partial_\mu \phi^\mu)(\partial_\nu \phi^\nu) + \frac{1}{2}(\partial_\mu \tilde \phi^\mu)(\partial_\nu \tilde \phi^\nu)\nonumber\\
&\equiv& {\cal L}_{(1)}  + \frac{1}{2}\, (\partial^\nu B_{\nu\mu})^2 
- \frac{1}{2}\,(\partial \cdot \phi)^2 + \frac{1}{2}\,(\partial \cdot \tilde \phi)^2.
\label{8}
\end{eqnarray}
In the above, different signs of the gauge-fixing terms have been chosen for the algebraic convenience and we have 
adopted the short-hand notations: $\partial_\mu \phi^\mu = (\partial \cdot \phi)$ and $\partial_\mu \tilde \phi^\mu 
= (\partial \cdot \tilde \phi)$.   At this stage, we note the following. First of all, we have the {\it discrete} 
symmetry transformations  in the theory because under the following transformations
\begin{eqnarray}
B_{\mu\nu} \to \mp\, \frac{i}{2}\, \varepsilon_{\mu\nu\lambda\rho}\, B^{\lambda\rho}, \qquad \phi_\mu \to \pm \,i \,\tilde \phi_\mu, 
\qquad \tilde \phi_\mu \to \mp \,i \,\phi_\mu,
\label{9}
\end{eqnarray}
the Lagrangian density ${\cal L}_{(2)}$ remains invariant (i.e. ${\cal L}_{(2)} \to {\cal L}_{(2)}$). 
This observation is interesting and important for us as its {\it generalized version}  (cf. Eqs. (\ref{14}), (\ref{22}), (\ref{82})) would play very important 
role in our proof of this model to be an example of the Hodge theory. Furthermore, we obtain the following 
EL-EOMs for the Lagrangian density ${\cal L}_{(2)}$
\begin{eqnarray}
&& (\Box + m^2)\, B_{\mu\nu} - m \,\Phi_{\mu\nu} - \frac{m}{2}\, \varepsilon_{\mu\nu\eta\kappa}\, \tilde \Phi^{\eta\kappa} = 0, \nonumber\\
&& \Box\, \phi_\mu - m \,(\partial^\nu B_{\nu\mu}) = 0, 
\qquad \Box \,\tilde \phi_\mu - \frac{m}{2}\, \varepsilon_{\mu\nu\eta\kappa}\, \partial^\nu B^{\eta\kappa} = 0, 
\label{10}
\end{eqnarray}
where the last equation can be $also $ written as $\Box \,\tilde \phi_\mu - \frac {m}{3!} \,\varepsilon_{\mu\nu\eta\kappa}\, H^{\nu\eta\kappa} = 0$. 
It can be {\it also} checked that we have: $\Box\, (\partial \cdot \phi) = 0$ and $\Box \,(\partial \cdot \tilde \phi) = 0$ 
from the last two equations of (\ref{10}) by applying a derivative on them.

The kinetic term (i.e. $\frac{1}{12}\, H_{\mu\nu\eta} H^{\mu\nu\eta}$) for the antisymmetric tensor ($B_{\mu\nu}$) field and 
gauge-fixing terms $[\frac{1}{2}\,(\partial^\nu B_{\mu\nu})^2 - \frac{1}{2}\, (\partial \cdot \phi)^2 
+ \frac{1}{2}\, (\partial \cdot \tilde \phi)^2]$ for the $B_{\mu\nu}$, $\phi_\mu$ and $\tilde \phi_\mu$ fields, respectively,  can be 
linearized by invoking the auxiliary fields $({\cal B}_\mu, B_\mu, B, {\cal B})$ as follows:
\begin{eqnarray}
{\cal L}_{(3)} &=& \frac{1}{2}{\cal B}_\mu {\cal B}^\mu - {\cal B}^\mu \left(\frac{1}{2}
\varepsilon_{\mu\nu\eta\kappa} \, \partial^\nu B^{\eta\kappa} \right) 
- \frac{m^2}{4} \,B^{\mu\nu}B_{\mu\nu}
- \frac{1}{4}\, \Phi^{\mu\nu}\Phi_{\mu\nu} \nonumber\\
&+& \frac{m}{2}\, B^{\mu\nu}\Phi_{\mu\nu} + \frac{1}{4}\, \tilde\Phi^{\mu\nu}\tilde\Phi_{\mu\nu} 
+ \frac{m}{4}\, \varepsilon^{\mu\nu\eta\kappa} B_{\mu\nu} \tilde\Phi_{\eta\kappa} 
- \frac{1}{2}\,B^\mu B_\mu \nonumber\\
&+& B^{\mu}\left(\partial^\nu B_{\nu\mu} \right) + \frac{1}{2}\, B^2 + B \left(\partial_\mu \phi^\mu \right)
- \frac{1}{2}\,{\cal B}^2 - {\cal B} \left(\partial_\mu \tilde \phi^\mu \right).
\label{11}
\end{eqnarray} 
At this stage, it is self-evident that the mass dimension of {\it all} the Nakanishi--Lautrup type auxiliary fields 
$({\cal B}_\mu, B_\mu, B, {\cal B})$ is $[M]^2$ for our {\it massive} 4D Abelian 2-form theory.  
It is straightforward to check that we have the following EL-EOMs 
w.r.t. the auxiliary fields $({\cal B}_\mu, B_\mu, {\cal B}, B)$:
\begin{eqnarray}
{\cal B}_\mu = \frac{1}{2}\, \varepsilon_{\mu\nu\eta\kappa}\,\partial^\nu B^{\eta\kappa}, 
\quad B_\mu = \partial^\nu B_{\nu\mu}, \quad{\cal B} = -(\partial \cdot \tilde \phi), \quad B = -(\partial \cdot  \phi).
\label{12}
\end{eqnarray}
We note that $(\partial \cdot B) = 0$, $(\partial \cdot {\cal B}) = 0$ and the substitution of 
these values of the auxiliary fields into ${\cal L}_{(3)}$ produces the Lagrangian density  ${\cal L}_{(2)}$. 
Furthermore, we note that $\Box \,{\cal B} = 0$ and $\Box\, B = 0$  because of $\Box\, (\partial \cdot \phi) = 0$ and 
$\Box\, (\partial \cdot \tilde \phi) = 0$.

The origin of these auxiliary fields is as follows. It is self-evident that, to linearize the 
gauge-fixing terms $[-\frac{1}{2}\, (\partial \cdot \phi)^2 + \frac{1}{2}\, (\partial \cdot \phi)^2 ]$ 
for the Abelian 1-form fields, we require the 0-form auxiliary fields $B$ and ${\cal B}$, respectively. However, for the linearization 
of the gauge-fixing term $[\frac{1}{2} (\partial^\nu B_{\nu\mu})^2]$ for the 2-form field, we require a 
1-form $B_\mu$ field. Since the kinetic term $(\frac{1}{12}\, H_{\mu\nu\eta} H^{\mu\nu\eta})$ corresponds 
to the Abelian  2-form field $B_{\mu\nu}$, we have to linearize it by using a 1-form which emerges from 
taking the Hodge dual of $H^{(3)}$ as:        
\begin{eqnarray}
*\, H^{(3)} = -\,\frac{1}{3!}\, dx^\mu \big(\varepsilon_{\mu\nu\eta\kappa}\, H^{\nu\eta\kappa} \big) 
\equiv  dx^\mu\, \Big(- \, \frac{1}{2}\, \varepsilon_{\mu\nu\eta\kappa}\, \partial^\nu B^{\eta\kappa} \Big). 
\label{13}
\end{eqnarray}
We have utilized the above expression in  linearizing the kinetic term for the Abelian 2-form gauge 
field $B_{\mu\nu}$ by taking the help of an auxiliary 1-form field 
${\cal B}_\mu$. The kinetic term for the 1-form fields $\phi_\mu$ and $\tilde \phi_\mu$ can {\it not} 
be linearized because we can {\it not} have a 0-form field to accomplish this goal.  Now the stage is set to 
discuss the {\it discrete} symmetries of the Lagrangian density ${\cal L}_{(3)}$. These are as follows:
\begin{eqnarray}
&& B_{\mu\nu} \to \mp \,\frac{i}{2}\, \varepsilon_{\mu\nu\eta\kappa} \, B^{\eta\kappa}, 
\qquad \phi_\mu \to \pm\, i\, \tilde \phi_\mu, \qquad  \tilde \phi_\mu \to \mp \,i\, \phi_\mu, \nonumber\\
&& B_\mu \to  \pm\, i \, {\cal B}_\mu, \qquad {\cal B}_\mu \to  \mp\, i \, B_\mu, \qquad  B \to \pm\, i\, {\cal B}, \qquad {\cal B} \to \mp\, i\, B.
\label{14}
\end{eqnarray}
We shall see later that these transformations (i.e. (\ref{14})) would play very important 
role within the framework of BRST formalism where their generalized forms (cf. Eqs. (\ref{22}), (\ref{82})) would be very useful.

We lay emphasis on the fact that the quantity $* H^{(3)} =  dx^\mu (-\frac{1}{2}\,\varepsilon_{\mu\nu\eta\kappa}\, \partial^\nu B^{\eta\kappa})$,
which has been used in the linearization of the kinetic term for the $B_{\mu\nu}$ field, is an axial-vector 1-form. Thus, there is a 
room for its generalization because we can always add/subtract an axial-vector field defined through an axial-vector 1-form 
$\tilde \varphi^{(1)}= dx^\mu \, \partial_\mu \tilde \varphi$ to it. Furthermore, an axial-vector of the kind 
$\tilde \phi^{(1)} = dx^\mu \,\tilde \phi_\mu$ can $also$ be added to it with proper {\it mass} dimension. Taking these inputs 
into account, we have  the following generalizations:
\begin{eqnarray}
&& \frac{1}{2}\, \varepsilon_{\mu\nu\eta\kappa} \partial^\nu B^{\eta\kappa} \to \frac{1}{2}\, \varepsilon_{\mu\nu\eta\kappa}\, \partial^\nu B^{\eta\kappa} 
- \frac{1}{2}\,  \partial_\mu \tilde \varphi + m \, \tilde \phi_\mu, \nonumber\\
&& \frac{1}{2}\, \varepsilon_{\mu\nu\eta\kappa}\, \partial^\nu B^{\eta\kappa} \to \frac{1}{2}\, \varepsilon_{\mu\nu\eta\kappa}\, \partial^\nu B^{\eta\kappa} 
+ \frac{1}{2}\,  \partial_\mu \tilde \varphi + m \,\tilde \phi_\mu. \qquad
\label{15}
\end{eqnarray} 
It should be noted that, in the above, there is a sign difference in the {\it second} term on the r.h.s.   
In exactly similar fashion, the gauge-fixing terms, which have been derived through the application of co-exterior 
derivative $\delta = - * d*$ (cf. (\ref{7})), can {\it also} be generalized as follows: 
\begin{eqnarray}
&& \partial^\nu B_{\nu\mu} \to \partial^\nu B_{\nu\mu} \mp \frac{1}{2}\, \partial_\mu \varphi + m \,\phi_\mu, \nonumber\\
&& \partial_\mu \phi^\mu \to \partial_\mu \phi^\mu \pm \frac{m}{2}\, \varphi, \qquad
\partial_\mu \tilde \phi^\mu \to \partial_\mu \tilde \phi^\mu \pm \frac{m}{2}\, \tilde \varphi.
\label{16}
\end{eqnarray}
It should be noted here that, because of the existence of the (pseudo)scalar $(\tilde \varphi) \varphi$ 
fields and (axial-)vector $(\tilde \phi_\mu) \phi_\mu$ fields in our theory, we have added/subtracted these fields 
with proper mass dimension.

With the above modifications, the {\it most} general form of the {\it coupled} Lagrangian densities 
(that would be useful for our further  discussions) are:
\begin{eqnarray}
{\cal L}_{(3)} \to {\cal L}^{(1)}_{(4)}  &=& \frac{1}{2}{\cal B}_\mu {\cal B}^\mu - {\cal B}^\mu \bigg(\frac{1}{2}
\varepsilon_{\mu\nu\eta\kappa} \, \partial^\nu B^{\eta\kappa} - \frac{1}{2}\,\partial_\mu \tilde \varphi + m \,\tilde \phi_\mu\bigg) \nonumber\\
&-& \frac{m^2}{4} \,B^{\mu\nu}B_{\mu\nu} - \frac{1}{4}\, \Phi^{\mu\nu}\Phi_{\mu\nu}  
+ \frac{m}{2}\, B^{\mu\nu}\Phi_{\mu\nu} + \frac{1}{4}\, \tilde\Phi^{\mu\nu}\tilde\Phi_{\mu\nu} \nonumber\\
&+& \frac{m}{4} \varepsilon^{\mu\nu\eta\kappa} B_{\mu\nu} \tilde\Phi_{\eta\kappa} 
- \frac{1}{2} B^\mu B_\mu 
+ B^{\mu}\left( \partial^\nu B_{\nu\mu} - \frac{1}{2} \partial_\mu \varphi + m \phi_\mu \right) \nonumber\\
&+& \frac{1}{2}\, B^2 + B \left(\partial_\mu \phi^\mu + \frac{m}{2} \,\varphi \right)
- \frac{1}{2}\,{\cal B}^2 - {\cal B} \left(\partial_\mu \tilde \phi^\mu + \frac{m}{2} \, \tilde\varphi \right), 
\label{17}
\end{eqnarray}
\begin{eqnarray}
{\cal L}_{(3)} \to {\cal L}^{(2)}_{(4)} &=& \frac{1}{2} \bar {\cal B}_\mu \bar {\cal B}^\mu + \bar {\cal B}^\mu \bigg(\frac{1}{2}
\varepsilon_{\mu\nu\eta\kappa} \, \partial^\nu B^{\eta\kappa} + \frac{1}{2}\,\partial_\mu \tilde \varphi + m \,\tilde \phi_\mu\bigg) \nonumber\\
&-& \frac{m^2}{4} \,B^{\mu\nu}B_{\mu\nu} - \frac{1}{4}\, \Phi^{\mu\nu}\Phi_{\mu\nu}  
+ \frac{m}{2}\, B^{\mu\nu}\Phi_{\mu\nu} + \frac{1}{4}\, \tilde\Phi^{\mu\nu}\tilde\Phi_{\mu\nu} \nonumber\\
&+& \frac{m}{4} \varepsilon^{\mu\nu\eta\kappa} B_{\mu\nu} \tilde\Phi_{\eta\kappa} 
- \frac{1}{2} \bar B^\mu \bar B_\mu 
- \bar B^{\mu}\left( \partial^\nu B_{\nu\mu} + \frac{1}{2}\, \partial_\mu \varphi + m\, \phi_\mu \right) \nonumber\\
&+& \frac{1}{2}\, \bar B^2 - \bar B \left(\partial_\mu \phi^\mu - \frac{m}{2} \,\varphi \right)
- \frac{1}{2}\,\bar {\cal B}^2 + \bar {\cal B} \left(\partial_\mu \tilde \phi^\mu - \frac{m}{2} \, \tilde\varphi \right).
\label{18}
\end{eqnarray}
Here we have invoked the Nakanishi--Lautrup type auxiliary fields $(\bar {\cal B}_\mu,\bar B_\mu, \bar B, \bar {\cal B})$ 
for the linearization of kinetic and gauge-fixing terms of our present theory.
It is straightforward to note that these auxiliary fields {\it also} have the mass dimension $[M]^2$ in 
the natural units. The above coupled Lagrangian densities lead to the following EL-EOM w.r.t. the auxiliary fields:
\begin{eqnarray}
&& {\cal B}_\mu = \frac{1}{2}\,\varepsilon_{\mu\nu\eta\kappa} \, \partial^\nu B^{\eta\kappa} 
- \frac{1}{2}\,\partial_\mu \tilde \varphi + m \,\tilde \phi_\mu, 
\qquad {\cal B} = - \Big(\partial \cdot \tilde \phi + \frac{m}{2}\, \tilde \varphi \Big),\nonumber\\
&& B_\mu = \partial^\nu B_{\nu\mu} - \frac{1}{2}\,\partial_\mu \varphi + m\, \phi_\mu, 
\qquad B = - \Big(\partial \cdot \phi + \frac{m}{2}\, \varphi \Big),\nonumber\\
&& \nonumber\\
&& \bar {\cal B}_\mu = - \Big(\frac{1}{2}\,
\varepsilon_{\mu\nu\eta\kappa} \, \partial^\nu B^{\eta\kappa} + \frac{1}{2}\,\partial_\mu \tilde \varphi + m\, \tilde \phi_\mu\Big), 
\qquad \bar {\cal B} = + \Big(\partial \cdot \tilde \phi - \frac{m}{2}\, \tilde \varphi \Big), \nonumber\\
&& \bar B_\mu = - \Big(\partial^\nu B_{\nu\mu} + \frac{1}{2}\,\partial_\mu \varphi + m \,\phi_\mu \Big), 
\qquad \bar B = + \Big(\partial \cdot \phi - \frac{m}{2}\, \varphi \Big). 
\label{19}
\end{eqnarray}
The above equations automatically lead to the following CF-type restrictions\footnote{We discuss the existence and 
emergence of the CF-type restrictions in a diagrammatic language in our Appendix C. We show that the {\it clustering} 
of fields at a point implies the existence of CF-type relations/conditions for our theory.}:
\begin{eqnarray}
&& B + \bar B + m\,\varphi = 0, \qquad {\cal B} + \bar {\cal B} + m\, \tilde\varphi = 0,\nonumber\\
&& B_\mu + \bar B_\mu + \partial_\mu\varphi = 0, \qquad {\cal B}_\mu + \bar {\cal B}_\mu + \partial_\mu  \tilde\varphi = 0.
\label{20}
\end{eqnarray}
The above conditions/constraints would play important roles in our discussions on the nilpotent (anti-)BRST and (anti-)co-BRST 
symmetries of the generalized versions of the coupled Lagrangian densities ${\cal L}^{(1)}_{(4)}$ and ${\cal L}^{(2)}_{(4)}$
where we shall include the Faddeev-Popov ghost terms (cf. Eqs. (\ref{28}), (\ref{29}) below). We would like to comment, 
in passing, that the following relations
\begin{eqnarray}
&& B - \bar B + 2\, \big(\partial \cdot \varphi \big) = 0, \qquad B_\mu - \bar B_\mu - 2\big(\partial^\nu B_{\nu\mu} + m \phi_\mu \big) = 0, 
\nonumber\\
&& {\cal B} - \bar {\cal B} +2\, \big(\partial \cdot \tilde\varphi \big) = 0,
\qquad {\cal B}_\mu - \bar {\cal B}_\mu - 2 \big(\varepsilon_{\mu\nu\eta\kappa}\, \partial^\nu B^{\eta\kappa} 
+ m \tilde\phi_\mu \big) = 0.\qquad
\label{21}
\end{eqnarray}
would also play some roles in our discussions. However, these would {\it not} be as 
important as the CF-type restrictions quoted in (\ref{20}). Furthermore, it may be pertinent to point out that the coupled 
(but equivalent) Lagrangian densities in (\ref{17}) and (\ref{18}) are the {\it most} general in the sense that they lead to the derivation of EL-EOM
(\ref{19}) which, ultimately, imply the CF-type restrictions (\ref{20}). 


We end our present section with the final brief remark on the existence of {\it discrete} symmetry transformations in 
our theory which is described by the coupled {\it but} equivalent Lagrangian densities ${\cal L}^{(1)}_{(4)}$ and 
${\cal L}^{(2)}_{(4)}$ (cf. Eqs. (\ref{28}) and (\ref{29})). It can be explicitly checked that under the following discrete transformations
\begin{eqnarray}
&&  \phi_\mu \to \pm\, i\, \tilde \phi_\mu, 
\qquad \tilde \phi_\mu \to \mp\, i\, \phi_\mu, \;\qquad \varphi \to \pm \,i\, \tilde \varphi, \;\qquad \tilde\varphi \to \mp\, i\, \varphi, \nonumber\\
&& B_\mu \to \pm \,i\, {\cal B}_\mu, \qquad {\cal B}_\mu \to \mp\, i\, B_\mu, 
\qquad B \to \pm\, i \,{\cal B}, \qquad {\cal B} \to \mp\, i\, B, \nonumber\\
&& \bar B_\mu \to \pm \,i\, \bar {\cal B}_\mu, \qquad \bar {\cal B}_\mu \to \mp \,i\,\bar  B_\mu, 
\qquad \bar B \to \pm\, i\, \bar {\cal B}, \qquad \bar {\cal B} \to \mp\, i\, \bar B, \nonumber\\
&& B_{\mu\nu}\to \mp\, \frac{i}{2}\, \varepsilon_{\mu\nu\eta\kappa} B^{\eta\kappa}, \qquad B_{\mu\nu} B^{\mu\nu} \to B_{\mu\nu} B^{\mu\nu},
\label{22}
\end{eqnarray}
the Lagrangian densities ${\cal L}^{(1)}_{(4)}$ and 
${\cal L}^{(2)}_{(4)}$ remain invariant. It is to be noted that the {\it mass} term of the Abelian 2-form gauge field 
(i.e. $-\,\frac{m^2}{4}\,\, B^{\mu\nu} B_{\mu\nu}$) remains invariant under the discrete symmetry transformation
($B_{\mu\nu}\to \mp\, \frac{i}{2}\, \varepsilon_{\mu\nu\eta\kappa} B^{\eta\kappa}$). Furthermore, we observe that the 
topological mass term (i.e. $\frac{m^2}{4}\,\varepsilon^{\mu\nu\eta\kappa}\, B_{\mu\nu} \tilde \Phi_{\eta\kappa}$)  
and mass term ($\frac{m}{2}\, B^{\mu\nu}\, \Phi_{\mu\nu}$) exchange with each other due to the discrete symmetry transformations (\ref{22}). 
Finally, we point out that the kinetic terms for $\phi_\mu$ and $\tilde \phi_\mu$ fields (i.e. $-\,\frac{1}{4}\, \Phi^{\mu\nu} \,\Phi_{\mu\nu}$ 
and $\frac{1}{4}\, \tilde \Phi^{\mu\nu}\, \tilde \Phi_{\mu\nu}$) exchange to each other due to symmetry transformations listed in (\ref{22}).  
These observations are exactly similar to the observations made in the context of 2D Proca theory (cf. Appendix A below).
We shall see  that the discrete symmetry transformations (\ref{22}) would be generalized within the framework of BRST formalism in 
Section \ref{Sect. 6} (see below) where the discrete symmetry transformations for the dynamical (anti-)ghost fields as well as auxiliary (anti-)ghost 
fields would also be incorporated (cf. Eq. (\ref{82}) below).


\section{Off-shell nilpotent (anti-)BRST symmetries} \label{Sect. 3}

We have discussed the ``classical" gauge symmetry transformations (\ref{5}) and (\ref{6}) in the previous section. These local gauge 
transformations can be generalized at the ``quantum" level, within the framework of BRST formalism, in the language  of the
continuous and infinitesimal (anti-)BRST symmetry transformations $s_{(a)b}$ as follows: 
\begin{eqnarray}
&&  s_{ab} B_{\mu\nu} = - (\partial_\mu \bar C_\nu - \partial_\nu \bar C_\mu), \qquad 
s_{ab} \bar C_\mu  = - \partial_\mu \bar \beta, \qquad s_{ab}  C_\mu =  \bar B_\mu, \nonumber\\
&& s_{ab} \beta = - \lambda, \quad s_{ab} \phi_\mu = \partial_\mu \bar C - m\, \bar C_\mu, \qquad s_{ab} B_\mu =  - \partial_\mu \rho,\nonumber\\
&& s_{ab} \bar C = - m\, \bar \beta, \qquad s_{ab}  C = \bar B, \qquad s_{ab} B = -\, m \,\rho, \qquad s_{ab} \varphi = \rho, \nonumber\\
&& s_{ab} [\bar B, \rho, \lambda, \bar \beta, \bar B_\mu, {\cal B}_\mu, {\bar{\cal B}}_\mu, 
\tilde\phi_\mu, \tilde \varphi, {\cal B}, \bar {\cal B},  H_{\mu\nu\kappa}] = 0,
\label{23} 
\end{eqnarray}
\begin{eqnarray}
&&  s_b B_{\mu\nu} = - (\partial_\mu C_\nu - \partial_\nu C_\mu), \qquad 
s_b C_\mu  = - \partial_\mu \beta, \qquad s_b \bar C_\mu = B_\mu, \nonumber\\
&& s_b \bar \beta = - \rho, \quad s_b \phi_\mu = \partial_\mu C - m\, C_\mu, \qquad s_b \bar B_\mu =  - \partial_\mu \lambda, \nonumber\\
&& s_b C = - \,m\,\beta, \qquad s_b \bar C =  B, \qquad s_b \bar B = -\, m\, \lambda,\qquad s_b \varphi = \lambda, \nonumber\\
&& s_b [B, \rho, \lambda, \beta, B_\mu, {\cal B}_\mu, {\bar {\cal B}}_\mu, \tilde\phi_\mu,
\tilde \varphi, {\cal B}, \bar {\cal B}, H_{\mu\nu\kappa}] = 0.
\label{24}
\end{eqnarray}
A few comments, at this stage, are in order. First of all, we note that the above
fermionic (anti-)BRST symmetry transformations $s_{(a)b}$ are off-shell nilpotent 
(i.e. $s_{(a)b}^2 = 0$) of order two. Second, it can be checked that the field 
strength tensor $H_{\mu\nu\eta}$ (owing its origin in the exterior derivative 
$d = dx^\mu \partial_\mu$) remains invariant under the (anti-)BRST transformations (i.e. $s_{(a)b}\, H_{\mu\nu\eta} = 0$). 
To be precise, we observe that {\it all } the fields, present in the kinetic term of the Abelian 2-form field 
$B_{\mu\nu}$ (cf. Eqs. (\ref{17}) and (\ref{18})), remain invariant under the (anti-)BRST symmetry transformations $s_{(a)b}$. 
Third, the nilpotent (anti-)BRST symmetry transformations are supersymmetric-type because they change bosonic fields into 
fermionic fields and {\it vice-versa}.  Fourth, we point out that the (anti-)BRST symmetry transformations $s_{(a)b}$ 
are absolutely anticommuting in nature
\begin{eqnarray}
&& \{s_b, \,s_{ab}\}\,B_{\mu\nu} = - \partial_\mu\big(B_\nu + \bar B_\nu \big) + \partial_\nu \big(B_\mu + \bar B_\mu \big), \nonumber\\
&& \{s_b, \,s_{ab}\}\,\phi_\mu = \partial_\mu\big(B + \bar B \big) - m\,\big(B_\mu + \bar B_\mu \big), 
\label{25}
\end{eqnarray}
provided we take into account the CF-type restrictions: $B_\mu + \bar B_\mu + \partial_\mu \varphi = 0$ and 
$B + \bar B + m\, \varphi = 0$ which have been derived in Eq. (\ref{20}). Finally, we note that the above CF-type restrictions are
(anti-)BRST invariant (i.e. $s_{(a)b} (B_\mu + \bar B_\mu + \partial_\mu \varphi) = 0$,
$s_{(a)b} (B +  \bar B + m \,\varphi) = 0$). As a consequence, these restrictions are ``physical" at the {\it quantum} level
which could be utilized, even from outside, for the specific proofs and purposes within the framework of BRST approach to our
present 4D $massive$ Abelian 2-form free gauge theory.

The above nilpotent and absolutely anticommuting (anti-)BRST transformations are the {\it symmetry} transformations for the
{\it specific} type of coupled (but equivalent) Lagrangian densities which are generalizations of the Lagrangian 
densities (\ref{17}) and (\ref{18}) as follows: 
\begin{eqnarray}
{\cal L}^{(1)}_{(4)} \to {\cal L}_{(B, {\cal B})} &=& {\cal L}^{(1)}_{(4)} 
+ s_b \,s_{ab} \bigg[-\frac{1}{2}\,\phi_\mu \phi^\mu + \frac{1}{4}\,B_{\mu\nu}B^{\mu\nu} - \frac{1}{2}\,\bar C^\mu C_\mu \nonumber\\
&-& \frac{1}{2}\,\bar C C + \frac{1}{4}\,\bar \beta \beta + \frac{1}{4}\, \varphi^2 
-  \frac{1}{4}\, \tilde \varphi^2 + \frac{1}{2}\, \tilde \phi^\mu \tilde \phi_\mu\bigg],
\label{26}
\end{eqnarray}
\begin{eqnarray}
{\cal L}^{(2)}_{(4)} \to {\cal L}_{(\bar B, \bar {\cal B})} &=& {\cal L}^{(2)}_{(4)} 
- s_{ab} \,s_b \bigg[-\frac{1}{2}\,\phi_\mu \phi^\mu + \frac{1}{4}\,B_{\mu\nu}B^{\mu\nu} - \frac{1}{2}\,\bar C^\mu C_\mu \nonumber\\
&-& \frac{1}{2}\,\bar C C  + \frac{1}{4}\,\bar \beta \beta + \frac{1}{4}\, \varphi^2 
-  \frac{1}{4}\, \tilde \varphi^2 + \frac{1}{2}\, \tilde \phi^\mu \tilde \phi_\mu\bigg],
\label{27}
\end{eqnarray}
where $s_{(a)b}$ are nothing but the (anti-)BRST symmetry transformations written in Eqs. (\ref{24}) and (\ref{23}). 
The above forms of Lagrangian densities imply, in a straightforward fashion, the BRST invariance of 
${\cal L}_{(B, {\cal B})}$ and anti-BRST invariance of ${\cal L}_{(\bar B, \bar {\cal B})}$ due to the
off-shell nilpotency (i.e. $s^2_{(a)b} = 0$) of   $s_{(a)b}$. As a consequence of the absolute anticommutativity 
of $s_{(a)b}$ (i.e. $s_b \, s_{ab} + s_{ab}\, s_b = 0$), it is also evident that the anti-BRST invariance of 
${\cal L}_{(B, {\cal B})}$ and BRST invariance of ${\cal L}_{(\bar B, \bar {\cal B})}$ would require the use 
of CF-type restrictions for their proof. This is due to the fact that the absolute anticommutativity 
($s_b \, s_{ab} + s_{ab}\, s_b = 0$) property of $s_{(a)b}$ is satisfied (if and only if the CF-type conditions are obeyed (cf. Eq. (\ref{25})).
To be more specific, it is clear that when $s_b$ would act on  ${\cal L}_{(\bar B, \bar {\cal B})}$, we have to use its
absolute anticommutativity property to prove the invariance of this specific Lagrangian density. Similar argument is
valid when $s_{ab}$ acts on  ${\cal L}_{(B, {\cal B})}$ to prove the anti-BRST invariance of {\it this} specific Lagrangian density
(i.e. ${\cal L}_{(B, {\cal B})}$).

It is interesting to mention here some of the specific features that are associated with the combination of 
fields that have been written in the parenthesis of Eqs. (\ref{26}) and (\ref{27}) on the r.h.s. We note, 
in this context, that the final ghost number of all the individual terms (in the parenthesis) is {\it zero} 
so that the application of $s_b$ and $s_{ab}$ {\it together} on these terms maintains this ghost number. 
In other words, the Lagrangian density should possess terms that carry the ghost number equal to zero. 
Furthermore, we observe that the mass dimension of all the individual terms is equal to {\it two} so that 
the applications of $s_b$ and $s_{ab}$ on the individual terms lead to the terms of the Lagrangian densities 
having the mass dimension {\it four} (as is required for a physically well-defined 4D theory which is renormalizable and consistent).

To corroborate the above statements, we derive here the explicit forms of the coupled (but equivalent)  Lagrangian densities 
so that we could apply the (anti-) BRST symmetry transformations $s_{(a)b}$ on {\it them} explicitly. The expanded and explicit
forms of these Lagrangian densities, in terms of the basic and auxiliary fields, are as follows: 
\begin{eqnarray}
{\cal L}_{(B,\cal B)} &=& \frac{1}{2}{\cal B}_\mu {\cal B}^\mu - {\cal B}^\mu \bigg(\frac{1}{2}
\varepsilon_{\mu\nu\eta\kappa} \, \partial^\nu B^{\eta\kappa} - \frac{1}{2}\,\partial_\mu \tilde \varphi + m \tilde \phi_\mu\bigg) 
- \frac{m^2}{4} \,B^{\mu\nu}B_{\mu\nu} \nonumber\\
&-& \frac{1}{4}\, \Phi^{\mu\nu}\Phi_{\mu\nu} + \frac{m}{2}\, B^{\mu\nu}\Phi_{\mu\nu} 
+ \frac{1}{4}\, \tilde\Phi^{\mu\nu}\tilde\Phi_{\mu\nu}
+ \frac{m}{4}\, \varepsilon^{\mu\nu\eta\kappa} B_{\mu\nu} \tilde\Phi_{\eta\kappa} \nonumber\\
&-& \frac{1}{2}\,B^\mu B_\mu  
+ B^{\mu}\left( \partial^\nu B_{\nu\mu} - \frac{1}{2}\, \partial_\mu \varphi + m \phi_\mu \right) + \frac{1}{2}\, B^2 \nonumber\\
&+& B \left(\partial_\mu \phi^\mu + \frac{m}{2} \,\varphi \right) - \frac{1}{2}\,{\cal B}^2 
- {\cal B} \Big(\partial_\mu \tilde \phi^\mu + \frac{m}{2} \, \tilde\varphi \Big) \nonumber\\
&+&  \big(\partial_\mu \bar C - m \bar C_\mu \big) \big(\partial^\mu C - m C^\mu \big) + \frac{1}{2}\, m^2\, \bar \beta \beta \nonumber\\
&-& \big(\partial_\mu \bar C_\nu - \partial_\nu \bar C_\mu \big) \big(\partial^\mu C^\nu \big) 
- \frac{1}{2}\,\partial_\mu \bar \beta \,\partial^\mu \beta  \nonumber\\
&-& \frac{1}{2}\left(\partial_\mu \bar C^\mu +  m \, \bar C + \frac{\rho}{4} \right) \lambda 
- \frac{1}{2}\left(\partial_\mu C^\mu +  m \, C - \frac{\lambda}{4} \right) \rho, 
\label{28}
\end{eqnarray}
\begin{eqnarray}
{\cal L}_{(\bar B,\bar {\cal B})} &=& \frac{1}{2} \bar {\cal B}_\mu \bar {\cal B}^\mu + \bar {\cal B}^\mu \bigg(\frac{1}{2}
\varepsilon_{\mu\nu\eta\kappa} \, \partial^\nu B^{\eta\kappa} + \frac{1}{2}\,\partial_\mu \tilde \varphi + m \tilde \phi_\mu\bigg) 
- \frac{m^2}{4} \,B^{\mu\nu}B_{\mu\nu} \nonumber\\
&-& \frac{1}{4}\, \Phi^{\mu\nu}\Phi_{\mu\nu} + \frac{m}{2}\, B^{\mu\nu}\Phi_{\mu\nu} 
+ \frac{1}{4}\, \tilde\Phi^{\mu\nu}\tilde\Phi_{\mu\nu}
+ \frac{m}{4}\, \varepsilon^{\mu\nu\eta\kappa} B_{\mu\nu} \tilde\Phi_{\eta\kappa} \nonumber\\
&-& \frac{1}{2}\, \bar B^\mu \bar B_\mu  
- \bar B^{\mu}\left( \partial^\nu B_{\nu\mu} + \frac{1}{2}\, \partial_\mu \varphi + m \phi_\mu \right) 
+ \frac{1}{2}\, \bar B^2 \nonumber\\
&-& \bar B \Big(\partial_\mu \phi^\mu - \frac{m}{2} \,\varphi \Big) 
- \frac{1}{2}\,\bar {\cal B}^2 + \bar {\cal B} \left(\partial_\mu \tilde \phi^\mu - \frac{m}{2} \, \tilde\varphi \right) \nonumber\\
&+&  \big(\partial_\mu \bar C - m \bar C_\mu \big) \big(\partial^\mu C - m C^\mu \big) + \frac{1}{2}\, m^2\, \bar \beta \beta \nonumber\\
&-& \big(\partial_\mu \bar C_\nu - \partial_\nu \bar C_\mu \big) \big(\partial^\mu C^\nu \big)
- \frac{1}{2}\,\partial_\mu \bar \beta \,\partial^\mu \beta  \nonumber\\
&-& \frac{1}{2}\left(\partial_\mu \bar C^\mu +  m \, \bar C + \frac{\rho}{4} \right) \lambda 
- \frac{1}{2}\left(\partial_\mu C^\mu +  m \, C - \frac{\lambda}{4} \right) \rho, 
\label{29}
\end{eqnarray}
where $(\bar C_\mu) C_\mu$ and $(\bar C)C$ are the fermionic (anti-)ghost ($\bar C^2_\mu = 0$, $C^2_\mu = 0$, $C_\mu C_\nu + C_\nu C_\mu = 0$, 
$\bar C_\mu \bar C_\nu + \bar C_\nu \bar C_\mu = 0$, $\bar C_\mu C_\nu + C_\nu \bar C_\mu = 0$, $C^2 = 0$, $\bar C^2 = 0$, 
$C \bar C + \bar C C = 0$, etc.)  fields which are the Lorentz vectors and scalars with ghost numbers $(-1)+1$, 
the bosonic (anti-)ghost fields $(\bar \beta) \beta$ carry the ghost number equal to $(-2) +2$, $(\rho) \lambda$ are the auxiliary 
(anti-)ghost fields with ghost numbers $(- 1) +1$, respectively. The rest of the symbols have already been explained in our previous 
section. Both the above Lagrangian densities are coupled because of the existence of the CF-type restrictions that are quoted in 
Eq. (\ref{20}). At this stage, it is  essential to mention that the mass dimension of $(\bar C_\mu, C_\mu, \bar \beta, \beta)$ is $[M]$ 
and that of $(\rho) \lambda$ is equal to $[M]^2$ (in natural units where $\hbar = c =1$).

The above coupled Lagrangian densities are {\it equivalent} on a submanifold of the field space where the 
CF-type restrictions (\ref{20}) are satisfied. This is due to the fact that {\it both} of them 
respect the (anti-)BRST symmetry transformations as
\begin{eqnarray}
s_b {\cal L}_{(B, {\cal B})} &=& - \partial_\mu \bigg[m\, \varepsilon^{\mu\nu\eta\kappa}\, \tilde \phi_\nu \big(\partial_\eta C_\kappa \big) 
+ B_\nu \big(\partial^\mu C^\nu - \partial^\nu C^\mu  \big) + \frac{1}{2}\, B^\mu\, \lambda \nonumber\\
&-& B \big(\partial^\mu C - m C^\mu\big) - \frac{1}{2}\, \big(\partial^\mu \beta \big)\,\rho \bigg],
\label{30}
\end{eqnarray}
\begin{eqnarray}
s_{ab} {\cal L}_{(\bar B, \bar {\cal B})} &=& - \partial_\mu \bigg[m\, \varepsilon^{\mu\nu\eta\kappa}\, 
\tilde \phi_\nu \big(\partial_\eta \bar C_\kappa \big) 
- \bar B_\nu \big(\partial^\mu \bar C^\nu - \partial^\nu \bar C^\mu  \big) + \frac{1}{2}\, \bar B^\mu\, \rho \nonumber\\
&+& \bar B \big(\partial^\mu \bar C - m \bar C^\mu\big) - \frac{1}{2}\, \big(\partial^\mu \bar \beta \big)\,\lambda \bigg],
\label{31}
\end{eqnarray}
\begin{eqnarray}
s_b {\cal L}_{(\bar B, \bar {\cal B})} &=& - \partial_\mu \bigg[m \varepsilon^{\mu\nu\eta\kappa} 
\tilde \phi_\nu \big(\partial_\eta C_\kappa \big) 
- \Big(\partial_\nu B^{\nu\mu} - \frac{1}{2} B^\mu + m \phi^\mu\Big) \lambda - \frac{1}{2} \big(\partial^\mu \beta \big)\,\rho \nonumber\\
&-& \bar B_\nu \big(\partial^\mu C^\nu - \partial^\nu C^\mu  \big)  
+ \bar B \big(\partial^\mu C - m  C^\mu\big)  \bigg] \nonumber\\
&+& \frac{1}{2}\, \big[B_\mu + \bar B_\mu + \partial_\mu \varphi \big] \big(\partial^\mu \lambda \big) 
- \partial_\mu\big[B_\nu + \bar B_\nu + \partial_\nu \varphi  \big] \big(\partial^\mu  C^\nu - \partial^\nu  C^\mu  \big) \nonumber\\
&-& m \big[B_\mu + \bar B_\mu + \partial_\mu \varphi \big] \big(\partial^\mu  C - m  C^\mu  \big)
- \frac{m}{2}\, \big[B + \bar B + m \varphi \big] \lambda \nonumber\\
&+& \partial_\mu \big[B + \bar B + m \varphi \big] \big(\partial^\mu  C - m  C^\mu  \big),
\label{32}
\end{eqnarray}
\begin{eqnarray}
s_{ab} {\cal L}_{(B, {\cal B})} &=& - \partial_\mu \bigg[m \varepsilon^{\mu\nu\eta\kappa} 
\tilde \phi_\nu \big(\partial_\eta \bar C_\kappa \big) 
+ \Big(\partial_\nu B^{\nu\mu} + \frac{1}{2} \bar B^\mu + m \phi^\mu\Big) \rho 
- \frac{1}{2} \big(\partial^\mu \bar \beta \big)\,\lambda \nonumber\\
&+& B_\nu \big(\partial^\mu \bar C^\nu - \partial^\nu \bar C^\mu  \big)  
- B \big(\partial^\mu \bar C - m \bar C^\mu\big) \bigg] \nonumber\\
&+& \frac{1}{2}\, \big[B_\mu + \bar B_\mu + \partial_\mu \varphi \big] \big(\partial^\mu \rho \big) 
+ \partial_\mu\big[B_\nu + \bar B_\nu + \partial_\nu \varphi  \big] \big(\partial^\mu \bar C^\nu - \partial^\nu \bar C^\mu  \big) \nonumber\\
&+& m \big[B_\mu + \bar B_\mu + \partial_\mu \varphi \big] \big(\partial^\mu \bar C - m \bar C^\mu  \big)
- \frac{m}{2}\, \big[B + \bar B + m \varphi \big] \rho \nonumber\\
&-& \partial_\mu \big[B + \bar B + m \varphi \big] \big(\partial^\mu \bar C - m \bar C^\mu  \big),
\label{33}
\end{eqnarray}
which demonstrate that, due to the validity of CF-type restrictions, we have:  
\begin{eqnarray}
s_b {\cal L}_{(\bar B, \bar {\cal B})} &=& - \partial_\mu \bigg[m\, \varepsilon^{\mu\nu\eta\kappa}\, 
\tilde \phi_\nu \big(\partial_\eta C_\kappa \big) - \frac{1}{2}\, \big(\partial^\mu \beta \big)\rho 
- \bar B_\nu \big(\partial^\mu C^\nu - \partial^\nu C^\mu  \big) \nonumber\\ 
&+& \bar B \big(\partial^\mu C - m  C^\mu\big)  
- \Big(\partial_\nu B^{\nu\mu} - \frac{1}{2}\, B^\mu + m \phi^\mu\Big) \lambda\,  \bigg],
\label{34}
\end{eqnarray}
\begin{eqnarray}
s_{ab} {\cal L}_{(B, {\cal B})} &=& - \partial_\mu \bigg[m\, \varepsilon^{\mu\nu\eta\kappa}\, 
\tilde \phi_\nu \big(\partial_\eta \bar C_\kappa \big) - \frac{1}{2}\, \big(\partial^\mu \bar \beta \big)\,\lambda 
+ B_\nu \big(\partial^\mu \bar C^\nu - \partial^\nu \bar C^\mu  \big) \nonumber\\  
&-& B \big(\partial^\mu \bar C - m \bar C^\mu\big) 
+ \Big(\partial_\nu B^{\nu\mu} + \frac{1}{2}\, \bar B^\mu + m \phi^\mu\Big) \rho \, \bigg]. 
\label{35} 
\end{eqnarray}
As a consequence, we note that {\it both} the action integrals $S_1 = \int d^4x\, {\cal L}_{(B, {\cal B})}$, 
$S_2 = \int d^4x\, {\cal L}_{(\bar B, \bar {\cal B})}$ respect {\it both} the off-shell nilpotent and absolutely anticommuting
symmetry transformations provided our whole theory is confined to be defined on a submanifold of the space of fields where 
the CF-type restrictions (\ref{20}) are satisfied.

According to the celebrated Noether theorem, the above invariances of the action integrals (w.r.t. the continuous and 
infinitesimal (anti-)BRST symmetry transformations) lead to the following Noether conserved currents: 
\begin{eqnarray}
J^\mu_{ab} &=& \varepsilon^{\mu\nu\eta\kappa}\,\big(m\, \tilde \phi_\nu + \bar {\cal B}_\nu \big) \big(\partial_\eta \bar C_\kappa \big)
+ \big(m B^{\mu\nu} - \Phi^{\mu\nu} \big) \big(\partial_\nu \bar C - m \bar C_\nu \big) \nonumber\\
&-& \bar B \big(\partial^\mu \bar C - m \bar C^\mu \big) - m \bar \beta \big(\partial^\mu C - m  C^\mu \big) 
+ \big(\partial^\mu C^\nu - \partial^\nu C^\mu \big) (\partial_\nu \bar \beta)\nonumber\\
&+& \bar B_\nu \big(\partial^\mu \bar C^\nu - \partial^\nu \bar C^\mu \big) + \frac{1}{2}\, (\partial^\mu \bar \beta)\, \lambda
- \frac{1}{2}\, \bar B^\mu\, \rho,
\label{36}
\end{eqnarray}
\begin{eqnarray}
J^\mu_b &=& \varepsilon^{\mu\nu\eta\kappa}\,\big(m\, \tilde \phi_\nu - {\cal B}_\nu \big) \big(\partial_\eta C_\kappa \big)
+ \big(m B^{\mu\nu} - \Phi^{\mu\nu} \big) \big(\partial_\nu C - m C_\nu \big) \nonumber\\
&+& B \big(\partial^\mu C - m C^\mu \big) + m \beta \big(\partial^\mu \bar C - m \bar C^\mu \big) 
- \big(\partial^\mu \bar C^\nu - \partial^\nu \bar C^\mu \big) (\partial_\nu \beta) \nonumber\\
&-& B_\nu \big(\partial^\mu C^\nu - \partial^\nu C^\mu \big) + \frac{1}{2}\, (\partial^\mu \beta)\, \rho
- \frac{1}{2}\, B^\mu\, \lambda.
\label{37}
\end{eqnarray}
The basic tenets of Noether's theorem enforce the condition that the above currents are  conserved on-shell. In other words,  
the conservation law (i.e. $\partial_\mu J^\mu_{(a)b} = 0$) can be proven by taking the help of the following EL-EOMs derived 
from ${\cal L}_{(B, {\cal B})}$, namely; 
\begin{eqnarray}
&& \varepsilon^{\mu\nu\eta\kappa}\, \partial_\mu {\cal B}_\nu + m^2\,\Big(B^{\eta\kappa} - \frac{1}{m}\, \Phi^{\eta\kappa}
- \frac{1}{2m}\, \varepsilon^{\mu\nu\eta\kappa}\, \tilde \Phi_{\mu\nu} \Big) 
+ \big(\partial^\eta B^\kappa - \partial^\kappa B^\eta \big)= 0, \nonumber\\
&& \varepsilon^{\mu\nu\eta\kappa}\, \partial_\mu B_\nu - \big(\partial^\eta {\cal B}^\kappa - \partial^\kappa {\cal B}^\eta \big) \nonumber\\
&&\, + \,\frac{m^2}{2}\,\varepsilon^{\mu\nu\eta\kappa}\,\Big(B_{\mu\nu} - \frac{1}{m}\, \Phi_{\mu\nu}
- \frac{1}{2m}\, \varepsilon_{\mu\nu\zeta\sigma}\, \tilde \Phi^{\zeta\sigma} \Big) = 0, \nonumber\\
&& \partial_\mu \Phi^{\mu\nu} - m \big(\partial_\mu B^{\mu\nu} - B^\nu \big) - \partial^\nu B = 0,  
\qquad B = - \left(\partial_\mu \phi^\mu + \frac{m}{2} \,\varphi \right),\nonumber\\
&& \partial_\mu \tilde \Phi^{\mu\nu} + m \Big(\frac{1}{2}\,\varepsilon^{\mu\nu\eta\kappa}\partial_\mu B_{\eta\kappa} 
+ {\cal B}^\nu \Big) - \partial^\nu {\cal B} = 0, 
\qquad {\cal B} = - \left(\partial_\mu \tilde \phi^\mu + \frac{m}{2}\, \tilde\varphi \right),  \nonumber\\
&& {\cal B}_\mu  = \left(\frac{1}{2}
\varepsilon_{\mu\nu\eta\kappa} \, \partial^\nu B^{\eta\kappa} - \frac{1}{2}\,\partial_\mu \tilde \varphi + m \tilde \phi_\mu\right), 
\qquad \partial_\mu {\cal B}^\mu + m {\cal B} = 0,\nonumber\\
&& B_{\mu} = \left(\partial^\nu B_{\nu\mu} - \frac{1}{2}\, \partial_\mu \varphi + m \phi_\mu \right), \qquad \partial_\mu B^\mu + m B = 0, \nonumber\\
&& \lambda = 2 \Big(\partial_\mu C^\mu + m C\Big), \qquad  \rho =  -2 \Big(\partial_\mu \bar C^\mu + m \bar C\Big), 
\qquad \big(\Box + m^2 \big) \beta = 0, \nonumber\\
&& \Box C - m \Big(\partial_\mu C^\mu - \frac{\lambda}{2} \Big) = 0, \quad 
\Box \bar C - m \Big(\partial_\mu \bar C^\mu + \frac{\rho}{2} \Big) = 0, \quad \big(\Box + m^2 \big) \bar\beta = 0,\nonumber\\
&& \big(\Box + m^2 \big) C_\mu  - \partial_\mu \Big(\partial_\nu C^\nu + m C - \frac{\lambda}{2} \Big)= 0, \nonumber\\ 
&& \Big(\Box + m^2 \Big) \bar C_\mu  - \partial_\mu\Big(\partial_\nu \bar C^\nu + m \bar C + \frac{\rho}{2} \Big)= 0,
\label{38}
\end{eqnarray}
and the EL-EOMs that are derived from ${\cal L}_{(\bar B, \bar {\cal B})}$ (and which are different from the above 
EL-EOMs from ${\cal L}_{(B,{\cal B})}$) are as follows:
\begin{eqnarray}
&& \varepsilon^{\mu\nu\eta\kappa}\, \partial_\mu \bar {\cal B}_\nu - m^2\,\Big(B^{\eta\kappa} - \frac{1}{m}\, \Phi^{\eta\kappa}
- \frac{1}{2m}\, \varepsilon^{\mu\nu\eta\kappa}\, \tilde \Phi_{\mu\nu} \Big)
+ \big(\partial^\eta \bar B^\kappa - \partial^\kappa \bar B^\eta \big)= 0, \nonumber\\
&& \varepsilon^{\mu\nu\eta\kappa}\, \partial_\mu \bar B_\nu  
- \big(\partial^\eta \bar {\cal B}^\kappa - \partial^\kappa \bar {\cal B}^\eta \big) \nonumber\\
&& \, -\, \frac{m^2}{2}\,\varepsilon^{\mu\nu\eta\kappa}\,\Big(B_{\mu\nu} - \frac{1}{m}\, \Phi_{\mu\nu}
- \frac{1}{2m}\, \varepsilon_{\mu\nu\zeta\sigma}\, \tilde \Phi^{\zeta\sigma} \Big) = 0, \nonumber\\
&& \partial_\mu \Phi^{\mu\nu} - m \big(\partial_\mu B^{\mu\nu} + \bar B^\nu \big) + \partial^\nu \bar B = 0, 
\qquad \bar B =  \left(\partial_\mu \phi^\mu - \frac{m}{2} \,\varphi \right), \nonumber\\ 
&& \partial_\mu \tilde \Phi^{\mu\nu} + m \Big(\frac{1}{2}\,\varepsilon^{\mu\nu\eta\kappa}\partial_\mu B_{\eta\kappa} 
- \bar {\cal B}^\nu \Big) + \partial^\nu \bar {\cal B} = 0, 
\qquad \bar {\cal B} =  \left(\partial_\mu \tilde \phi^\mu - \frac{m}{2}\, \tilde\varphi \right), \nonumber\\
&& \bar {\cal B}_\mu  = - \left(\frac{1}{2}
\varepsilon_{\mu\nu\eta\kappa} \, \partial^\nu B^{\eta\kappa} + \frac{1}{2}\,\partial_\mu \tilde \varphi + m \tilde \phi_\mu\right), 
\qquad \partial_\mu \bar {\cal B}^\mu + m \bar {\cal B} = 0, \nonumber\\
&& \bar B_{\mu} = - \left(\partial^\nu B_{\nu\mu} + \frac{1}{2}\, \partial_\mu \varphi + m \phi_\mu \right), 
\qquad \partial_\mu \bar B^\mu + m \bar B = 0. 
\label{39}
\end{eqnarray}
The zero component of the above currents in (\ref{36}) and (\ref{37}) leads to the definition of conserved Noether charges according to 
the Noether theorem. The (anti-) BRST charges $Q_{(a)b} = \int d^3x\, J^0_{(a)b}$ can be readily calculated from $J^\mu_{(a)b}$ 
(with $\varepsilon^{0ijk} = \epsilon^{ijk} \equiv -\,\epsilon_{ijk}$) as: 
\begin{eqnarray}
Q_{ab} &=& \int d^3x \Big[\epsilon^{ijk}\,\big(m\, \tilde \phi_i + \bar {\cal B}_i \big) \big(\partial_j \bar C_k \big)
+ \big(m B^{0i} - \Phi^{0i} \big) \big(\partial_i \bar C - m \bar C_i \big) \nonumber\\
&-& \bar B \big(\partial^0 \bar C - m \bar C^0 \big) - m \bar \beta \big(\partial^0 C - m  C^0 \big) 
+ (\partial_i \bar \beta) \big(\partial^0 C^i - \partial^i C^0 \big) \nonumber\\
&+& \bar B_i \big(\partial^0 \bar C^i - \partial^i \bar C^0 \big) + \frac{1}{2}\, (\partial^0 \bar \beta)\, \lambda
- \frac{1}{2}\, \bar B^0\, \rho \Big],
\label{40}
\end{eqnarray}
\begin{eqnarray}
Q_b &=& \int d^3x \Big[\epsilon^{ijk}\,\big(m\, \tilde \phi_i - {\cal B}_i \big) \big(\partial_j C_k \big)
+ \big(m B^{0i} - \Phi^{0i} \big) \big(\partial_i C - m C_i \big) \nonumber\\
&+& B \big(\partial^0 C - m C^0 \big) + m \beta \big(\partial^0 \bar C - m \bar C^0 \big) 
- (\partial_i \beta) \big(\partial^0 \bar C^i - \partial^i \bar C^0 \big)  \nonumber\\
&-& B_i \big(\partial^0 C^i - \partial^i C^0 \big) + \frac{1}{2}\, (\partial^0 \beta)\, \rho
- \frac{1}{2}\, B^0\, \lambda \Big].
\label{41}
\end{eqnarray}
The above charges are the generators for the continuous (anti-)BRST symmetry transformations as we have the following
\begin{eqnarray}
s_r \Psi = \pm\, i\, \big[\Psi, \, Q_r \big]_{(\pm)}, \qquad r = b, ab,
\label{42}
\end{eqnarray}
where $(\pm)$ signs, as the subscripts on the square bracket, denote the bracket to be the (anti)commutator for the 
generic field $\Psi$ being (fermionic)bosonic in nature. The decisive feature of the (anti-)BRST symmetry transformation 
is the observation that the curvature  (i.e. the field strength) tensor $H_{\mu\nu\eta}$, owing its origin to the 
exterior derivative (i.e. $d B^{(2)} = H^{(3)} = \frac{1}{3}\, (dx^\mu \wedge dx^\nu \wedge dx^\eta)\, H_{\mu\nu\eta}$), 
remains invariant under them (cf. Eqs. (\ref{23}), (\ref{24})).

We end this section with the final remark that the  nilpotency ($Q^2_{(a)b} = 0$) of the conserved (anti-)BRST 
charges can be proven by using the general relationship (\ref{42}), namely; 
\begin{eqnarray}
s_b\, Q_b = - i \, \big\{Q_b, \, Q_b \big\} = 0\; \Rightarrow\;  Q^2_b = 0, \nonumber\\
s_{ab}\, Q_{ab} =  - i \, \big\{Q_{ab}, \, Q_{ab} \big\} = 0 \;\Rightarrow  \;Q^2_{ab} = 0,
\label{43} 
\end{eqnarray}
where the l.h.s. can be computed precisely by using directly Eqs. (\ref{24}), (\ref{41}) and Eqs. (\ref{23}), (\ref{40}) for 
the clinching proof of (\ref{43}). The above Eq. (\ref{43}) has been written for the continuous symmetries $s_{(a)b}$ 
which are generated by the conserved and nilpotent (anti-)BRST charges $Q_{(a)b}$.


\section{Off-shell nilpotent (anti-)co-BRST symmetries} \label{Sect. 4}

The (anti-)BRST invariant Lagrangian densities ${\cal L}_{(B, {\cal B})}$ and
${\cal L}_{(\bar B, \bar {\cal B})}$ are {\it also} endowed with a set of fermionic (i.e. nilpotent) dual-BRST  
(i.e. co-BRST) and anti-dual (i.e. anti-co-BRST) symmetry transformations $s_{(a)d}$ as 
\begin{eqnarray}
&&  s_{ad} B_{\mu\nu} = - \varepsilon_{\mu\nu\eta\kappa} \,\partial^\eta  C^\kappa, \qquad 
s_{ad} \bar C_\mu = {\bar{\cal  B}}_\mu, \qquad s_{ad} C_\mu  =  \partial_\mu  \beta, \nonumber\\
&& s_{ad}  \bar\beta =  \rho, \quad  s_{ad} \tilde\phi_\mu = \partial_\mu  C - m\,  C_\mu, 
\qquad s_{ad} {\cal{ B}}_\mu =   \partial_\mu \lambda,\nonumber\\
&&  s_{ad} {\bar C} =  \bar {\cal  B}, \qquad  \quad s_{ad}   C =  m\, \beta, 
 \qquad s_{ad}  {\cal B} =  m \,\lambda, \qquad s_{ad} \tilde \varphi = -\,\lambda, \nonumber\\
&& s_{ad} [(\partial^\nu B_{\nu\mu}), B_\mu, {\cal{\bar B}}_\mu,
 \bar B_\mu, \bar {\cal B}, B, \bar B, \varphi, \phi_\mu, \rho, \lambda, \beta] = 0,\qquad
 \label{44}
\end{eqnarray}
\begin{eqnarray}
&&  s_d B_{\mu\nu} = - \varepsilon_{\mu\nu\eta\kappa} \,\partial^\eta \bar C^\kappa, \qquad 
s_d  C_\mu = {\cal B}_\mu, \qquad s_d \bar C_\mu  = - \partial_\mu \bar \beta, \nonumber\\
&& s_d  \beta = - \lambda, \qquad s_d \tilde\phi_\mu = \partial_\mu \bar C - m\, \bar C_\mu, 
\qquad s_d {\bar {\cal B}}_\mu =   \partial_\mu \rho, \nonumber\\
&& s_d  C =  {\cal B}, \qquad  \quad s_d \bar  C = - \,m\,\bar \beta, 
 \qquad s_d \bar {\cal B} =  m\, \rho, \qquad s_d \tilde \varphi = -\,\rho, \nonumber\\
&& s_d [(\partial^\nu B_{\nu\mu}), B_\mu, {\cal{ B}}_\mu,
 \bar B_\mu, {\cal B}, B, \bar B, \varphi, \phi_\mu, \rho, \lambda, \bar\beta] = 0, \qquad
 \label{45}
\end{eqnarray}
because  the Lagrangian densities ${\cal L}_{(B, {\cal B})}$ and
${\cal L}_{(\bar B, \bar {\cal B})}$ transform, under the above continuous and infinitesimal (anti-)co-BRST transformations, as 
\begin{eqnarray}
s_{ad} {\cal L}_{(\bar B, {\cal{\bar B}})} &=& - \partial_\mu \, 
\bigg[ m\, \varepsilon^{\mu\nu\eta\kappa} \phi_\nu \big(\partial_\eta C_\kappa \big) 
+ \bar {\cal B}_\nu \big(\partial^\mu C^\nu-\partial^\nu C^\mu \big) + \frac{1}{2}\,\bar {\cal B}^\mu\, \lambda  \nonumber\\
&-& \bar {\cal B} \big(\partial^\mu C - m \,C^\mu \big) + \frac{1}{2}\, \big(\partial^\mu \beta \big)\,\rho \bigg], 
\label{46}
\end{eqnarray}
\begin{eqnarray}
s_d {\cal L}_{(B, {\cal B})} &=& - \partial_\mu \, 
\bigg[ m\, \varepsilon^{\mu\nu\eta\kappa} \phi_\nu \big(\partial_\eta \bar C_\kappa \big) 
- {\cal B}_\nu \big(\partial^\mu {\bar C}^\nu-\partial^\nu \bar C^\mu \big) + \frac{1}{2}\, {\cal B}^\mu\, \rho  \nonumber\\
&+&  {\cal B}\, \big(\partial^\mu \bar C - m \,\bar C^\mu \big) - \frac{1}{2}\, \big(\partial^\mu \bar\beta \big)\,\lambda \bigg], 
\label{47}
\end{eqnarray}
which demonstrate that the action integrals corresponding to 
${\cal L}_{(B, {\cal B})}$ and ${\cal L}_{(\bar B, \bar {\cal B})}$: $S_1 = \int d^4x \, {\cal L}_{(B, {\cal B})}$ and 
$S_2 = \int d^4x \, {\cal L}_{(\bar B, \bar {\cal B})}$  remain invariant under the (anti-) co-BRST 
symmetry transformations for the physical  fields that vanish-off at $x \to\, \pm\, \infty$. Thus, we observe
that the Lagrangian density  ${\cal L}_{(B, {\cal B}}) $ respects the nilpotent co-BRST symmetry in a {\it perfect} manner as is the case
with the Lagrangian density ${\cal L}_{(\bar B, \bar {\cal B}})$ under the nilpotent anti-co-BRST transformations.

The above symmetry invariance happens because we have to, first of all, find out the consequences of the
application of $s_d$ and $s_{ad}$ on the combinations of fields that are present in the parenthesis of 
Eqs. (\ref{26}) and (\ref{27}) on the r.h.s. In this context, we note the following very useful and 
interesting observations 
\begin{eqnarray}
&& s_d \,s_{ad} \Big[-\frac{1}{2}\,\phi_\mu \phi^\mu + \frac{1}{4}\,B_{\mu\nu}B^{\mu\nu} - \frac{1}{2}\,\bar C^\mu C_\mu
- \frac{1}{2}\,\bar C C +  \frac{1}{4}\,\bar \beta \beta + \frac{1}{4}\, \varphi^2  \nonumber\\
&& \qquad \qquad \qquad -\,  \frac{1}{4}\, \tilde \varphi^2 
+ \frac{1}{2} \,\tilde \phi^\mu \tilde \phi_\mu\Big]\nonumber\\
&& \quad =\, \frac{1}{2}{\cal B}_\mu {\cal B}^\mu - {\cal B}^\mu \left(\frac{1}{2}
\varepsilon_{\mu\nu\eta\kappa} \, \partial^\nu B^{\eta\kappa} - \frac{1}{2}\,\partial_\mu \tilde \varphi + m \tilde \phi_\mu\right) \nonumber\\
&& \quad -\, \frac{1}{2}\,{\cal B}^2 - {\cal B} \left(\partial_\mu \tilde \phi^\mu + \frac{m}{2} \, \tilde\varphi \right)  
+ \frac{1}{2}\, m^2\, \bar \beta \beta \nonumber\\ 
&& \quad + \, \big(\partial_\mu \bar C - m \bar C_\mu \big) \big(\partial^\mu C - m C^\mu \big) 
- \frac{1}{2}\,\partial_\mu \bar \beta \,\partial^\mu \beta \nonumber\\
&& \quad -\, \big(\partial_\mu \bar C_\nu - \partial_\nu \bar C_\mu \big) \big(\partial^\mu C^\nu \big)
 - \frac{1}{2}\Big(\partial_\mu \bar C^\mu +  m \, \bar C + \frac{\rho}{4} \Big) \lambda \nonumber\\
&& \quad -\, \frac{1}{2}\left(\partial_\mu C^\mu +  m \, C - \frac{\lambda}{4} \right) \rho, 
\label{48}
\end{eqnarray}
and
\begin{eqnarray}
&& -\, s_{ad} \,s_d \Big[-\frac{1}{2}\,\phi_\mu \phi^\mu + \frac{1}{4}\,B_{\mu\nu}B^{\mu\nu} - \frac{1}{2}\,\bar C^\mu C_\mu
- \frac{1}{2}\,\bar C C + \frac{1}{4}\,\bar \beta \beta + \frac{1}{4}\, \varphi^2 \nonumber\\
&& \qquad \qquad \qquad  - \, \frac{1}{4}\, \tilde \varphi^2 
+ \frac{1}{2}\, \tilde \phi^\mu \tilde \phi_\mu\Big]\nonumber\\
&& \quad = \, \frac{1}{2} \bar {\cal B}_\mu \bar {\cal B}^\mu + \bar {\cal B}^\mu \left(\frac{1}{2}
\varepsilon_{\mu\nu\eta\kappa} \, \partial^\nu B^{\eta\kappa} + \frac{1}{2}\,\partial_\mu \tilde \varphi + m \tilde \phi_\mu\right) \nonumber\\ 
&& \quad -\, \frac{1}{2}\,\bar {\cal B}^2 + \bar {\cal B} \left(\partial_\mu \tilde \phi^\mu - \frac{m}{2} \, \tilde\varphi \right)
+ \frac{1}{2}\, m^2\, \bar \beta \beta \nonumber\\ 
&& \quad +\,  \big(\partial_\mu \bar C - m \bar C_\mu \big) \big(\partial^\mu C - m C^\mu \big) 
- \frac{1}{2}\,\partial_\mu \bar \beta \,\partial^\mu \beta \nonumber\\
&&\quad -\, \big(\partial_\mu \bar C_\nu - \partial_\nu \bar C_\mu \big) \big(\partial^\mu C^\nu \big)
- \frac{1}{2}\Big(\partial_\mu \bar C^\mu +  m \, \bar C + \frac{\rho}{4} \Big) \lambda \nonumber\\
&& \quad -\, \frac{1}{2}\left(\partial_\mu C^\mu +  m \, C - \frac{\lambda}{4} \right) \rho, 
\label{49}
\end{eqnarray}
which are nothing but the sum of the kinetic term for $B_{\mu\nu}$ field, gauge-fixing term for the axial-vector field and 
the Faddeev-Popov ghost terms. As a consequence of the above observations, we can write the Lagrangian 
densities ${\cal L}_{(B, {\cal B})}$ and ${\cal L}_{(\bar B, \bar {\cal B})}$,  in their expanded and explicit forms,
 as follows:    
\begin{eqnarray}
{\cal L}_{(B, {\cal B})} &=& \frac{m}{2}\, B^{\mu\nu}\Phi_{\mu\nu} + \frac{1}{4}\, \tilde\Phi^{\mu\nu}\tilde\Phi_{\mu\nu}
+ \frac{m}{4}\, \varepsilon^{\mu\nu\eta\kappa} B_{\mu\nu} \tilde\Phi_{\eta\kappa} - \frac{m^2}{4} \,B^{\mu\nu}B_{\mu\nu} \nonumber\\
&-& \frac{1}{4}\, \Phi^{\mu\nu}\Phi_{\mu\nu} 
- \frac{1}{2}\,B^\mu B_\mu + B^{\mu}\left( \partial^\nu B_{\nu\mu} - \frac{1}{2}\, \partial_\mu \varphi + m \phi_\mu \right) \nonumber\\
&+& \frac{1}{2}\, B^2 + B \left(\partial_\mu \phi^\mu + \frac{m}{2} \,\varphi \right) \nonumber\\
&+& s_d\, s_{ad} \Big[-\frac{1}{2}\,\phi_\mu \phi^\mu + \frac{1}{4}\,B_{\mu\nu}B^{\mu\nu} - \frac{1}{2}\,\bar C^\mu C_\mu 
- \frac{1}{2}\,\bar C C + \frac{1}{4}\,\bar \beta \beta \nonumber\\
&+& \frac{1}{4}\, \varphi^2 -  \frac{1}{4}\, \tilde \varphi^2 
+ \frac{1}{2}\, \tilde \phi^\mu \tilde \phi_\mu\Big], 
\label{50}
\end{eqnarray}
\begin{eqnarray}
{\cal L}_{(\bar B, \bar {\cal B})} &=& \frac{m}{2}\, B^{\mu\nu}\Phi_{\mu\nu} + \frac{1}{4}\, \tilde\Phi^{\mu\nu}\tilde\Phi_{\mu\nu}
+ \frac{m}{4}\, \varepsilon^{\mu\nu\eta\kappa} B_{\mu\nu} \tilde\Phi_{\eta\kappa} - \frac{m^2}{4} \,B^{\mu\nu}B_{\mu\nu} \nonumber\\
&-& \frac{1}{4}\, \Phi^{\mu\nu}\Phi_{\mu\nu}  
- \frac{1}{2}\, \bar B^\mu \bar B_\mu - \bar B^{\mu}\left( \partial^\nu B_{\nu\mu} + \frac{1}{2}\, \partial_\mu \varphi + m \phi_\mu \right) \nonumber\\
&+& \frac{1}{2}\, \bar B^2 - \bar B \left(\partial_\mu \phi^\mu - \frac{m}{2} \,\varphi \right) \nonumber\\
&-& s_{ad} \,s_d \Big[-\frac{1}{2}\,\phi_\mu \phi^\mu + \frac{1}{4}\,B_{\mu\nu}B^{\mu\nu} - \frac{1}{2}\,\bar C^\mu C_\mu 
- \frac{1}{2}\,\bar C C + \frac{1}{4}\,\bar \beta \beta \nonumber\\
&+& \frac{1}{4}\, \varphi^2 -  \frac{1}{4}\, \tilde \varphi^2 
+ \frac{1}{2}\, \tilde \phi^\mu \tilde \phi_\mu\Big].
\label{51}
\end{eqnarray}
The above mathematical expressions prove the dual-BRST (i.e. co-BRST) invariance of ${\cal L}_{(B, {\cal B})}$  and anti-dual-BRST
(i.e. anti-co-BRST) invariance 
of ${\cal L}_{(\bar B, \bar {\cal B})}$ due to the off-shell nilpotency (i.e. $s^2_{(a)d} = 0$) of the (anti-)co-BRST 
transformations $(s_{(a)d})$ that are present in our theory. In other words, we have:
\begin{eqnarray}
s_d {\cal L}_{(\bar B, \bar {\cal B})} &=& - \partial_\mu \Big[m\, \varepsilon^{\mu\nu\eta\kappa}\,  
\phi_\nu \big(\partial_\eta \bar C_\kappa \big) \Big],
\label{52}
\end{eqnarray}
\begin{eqnarray}
s_{ad} {\cal L}_{({\cal B},  {\cal B})} &=& - \partial_\mu \Big[m\, \varepsilon^{\mu\nu\eta\kappa}\, 
\phi_\nu \big(\partial_\eta C_\kappa \big) \Big].
\label{53}
\end{eqnarray}
The stage is now set to discuss  the absolute anticommutativity of the (anti-)co-BRST symmetry  transformations.
In this context,  we observe: 
\begin{eqnarray} 
&&{\{s_d, s_{ad}\}}\,B_{\mu\nu} = -\,\varepsilon_{\mu\nu\eta\kappa}\partial^\eta({\cal B}^{\kappa} + \bar {\cal B}^\kappa),\nonumber\\
&&{\{s_d, s_{ad}\}}\, \tilde\phi_\mu = \partial_\mu({\cal B} + \bar {\cal B}) - m\,({\cal B}_\mu + \bar {\cal B}_\mu).
\label{54}
\end{eqnarray}
It is straightforward to note that, for the absolute anticommutativity property (i.e. ${\{s_d, s_{ad}\}} = 0$) 
to be true, we have to invoke the CF-type restrictions: ${\cal B} + \bar {\cal B} + m\,\tilde \varphi  = 0$ and
${\cal B}_\mu + \bar {\cal B}_\mu + \partial_\mu\tilde\varphi = 0$.
We draw the conclusion that the property of the absolute anticommutativity is valid if and only if the 
CF-type restrictions are satisfied. The key consequence of the above result is the observation that the Lagrangian density
${\cal L}_{(B,\,  {\cal B})}$  respects the anti-dual BRST symmetry, too, provided we invoke the potential and
power of the CF-type restrictions. In exactly similar fashion, we note that the Lagrangian density ${\cal L}_{{(\bar  B},\, \bar{\cal B})}$
respects the dual-BRST symmetry transformations ($s_d$) if we confine ourselves to a submanifold 
in the space of fields where the CF-type restrictions are satisfied. Mathematically, we observe the following 
\begin{eqnarray}
s_d {\cal L}_{(\bar B, \bar {\cal B})} &=& - \partial_\mu \bigg[m\, \varepsilon^{\mu\nu\eta\kappa}\,  
\phi_\nu \big(\partial_\eta \bar C_\kappa \big) - \bar {\cal B} \big(\partial^\mu \bar C - m \bar C^\mu\big) 
- \frac{1}{2}\, \big(\partial^\mu \bar \beta \big)\,\lambda \nonumber\\
&+& \bar {\cal B}_\nu \big(\partial^\mu \bar C^\nu - \partial^\nu \bar C^\mu  \big)  - \Big(\frac{1}{2}\,\varepsilon^{\mu\nu\eta\kappa}\,\partial_\nu B_{\eta\kappa} 
- \frac{1}{2}\, {\cal B}^\mu + m \tilde \phi^\mu\Big) \rho  \bigg] \nonumber\\
&+& \frac{1}{2} \big[{\cal B}_\mu + \bar {\cal B}_\mu + \partial_\mu \tilde \varphi \big] \big(\partial^\mu \rho \big) 
- \frac{m}{2} \big[{\cal B} + \bar {\cal B} + m \tilde \varphi \big] \rho \nonumber\\
&+& \partial_\mu\big[{\cal B}_\nu + \bar {\cal B}_\nu + \partial_\nu \tilde \varphi  \big] \big(\partial^\mu \bar C^\nu 
- \partial^\nu \bar C^\mu  \big) \nonumber\\
&+& m \big[{\cal B}_\mu + \bar {\cal B}_\mu + \partial_\mu \tilde \varphi \big] \big(\partial^\mu \bar C - m \bar C^\mu  \big) \nonumber\\
&-& \partial_\mu \big[{\cal B} + \bar {\cal B} + m \tilde \varphi \big] \big(\partial^\mu \bar C - m \bar C^\mu  \big)
\label{55}
\end{eqnarray}
\begin{eqnarray}
s_{ad} {\cal L}_{( B,  {\cal B})} &=& - \partial_\mu \bigg[m\, \varepsilon^{\mu\nu\eta\kappa}\, 
\phi_\nu \big(\partial_\eta C_\kappa \big) + {\cal B} \big(\partial^\mu C - m  C^\mu\big) 
+ \frac{1}{2}\, \big(\partial^\mu \beta \big)\,\rho \nonumber\\
&+& \Big(\frac{1}{2}\,\varepsilon^{\mu\nu\eta\kappa}\,\partial_\nu B_{\eta\kappa} + \frac{1}{2}\, \bar {\cal B}^\mu 
+ m \tilde \phi^\mu\Big) \lambda - {\cal B}_\nu \big(\partial^\mu C^\nu - \partial^\nu C^\mu  \big) \bigg] \nonumber\\
&+& \frac{1}{2}\, \big[{\cal B}_\mu + \bar {\cal B}_\mu + \partial_\mu \tilde \varphi \big] \big(\partial^\mu \lambda \big) 
- \frac{m}{2}\, \big[{\cal B} + \bar {\cal B} + m \tilde \varphi \big] \lambda \nonumber\\
&-& \partial_\mu\big[{\cal B}_\nu + \bar {\cal B}_\nu + \partial_\nu \tilde \varphi  \big] \big(\partial^\mu  C^\nu - \partial^\nu  C^\mu  \big) \nonumber\\
&-& m \big[{\cal B}_\mu + \bar {\cal B}_\mu + \partial_\mu \tilde \varphi \big] \big(\partial^\mu  C - m  C^\mu  \big) \nonumber\\
&+& \partial_\mu \big[{\cal B} + \bar {\cal B} + m \tilde \varphi \big] \big(\partial^\mu  C - m  C^\mu  \big),
\label{56}
\end{eqnarray}
which capture the sanctity of the {\it statements} made in the paragraph above these equations. In other words, if we 
imposes the CF-type restrictions from {\it outside}, we obtain the following transformations for 
${\cal L}_{(B,{\cal B})}$ and ${\cal L}_{(\bar B, \bar {\cal B})}$
\begin{eqnarray}
s_d {\cal L}_{(\bar B, \bar {\cal B})} &=& - \partial_\mu \bigg[m \varepsilon^{\mu\nu\eta\kappa}  
\phi_\nu \big(\partial_\eta \bar C_\kappa \big) - \bar {\cal B} \big(\partial^\mu \bar C - m \bar C^\mu\big) 
- \frac{1}{2}  \big(\partial^\mu \bar \beta \big) \lambda  \nonumber\\
&-& \Big(\frac{1}{2} \varepsilon^{\mu\nu\eta\kappa} \partial_\nu B_{\eta\kappa} 
- \frac{1}{2}  {\cal B}^\mu + m \tilde \phi^\mu\Big) \rho 
+ \bar {\cal B}_\nu \big(\partial^\mu \bar C^\nu - \partial^\nu \bar C^\mu  \big) \bigg], 
\label{57}
\end{eqnarray}
\begin{eqnarray}
s_{ad} {\cal L}_{( B,  {\cal B})} &=& - \partial_\mu \bigg[m \varepsilon^{\mu\nu\eta\kappa}
\phi_\nu \big(\partial_\eta C_\kappa \big) + {\cal B} \big(\partial^\mu C - m  C^\mu\big) 
+ \frac{1}{2} \big(\partial^\mu \beta \big) \rho  \nonumber\\
&+& \Big(\frac{1}{2} \varepsilon^{\mu\nu\eta\kappa} \partial_\nu B_{\eta\kappa}
 + \frac{1}{2} \bar {\cal B}^\mu + m \tilde \phi^\mu\Big) \lambda 
-  {\cal B}_\nu \big(\partial^\mu C^\nu - \partial^\nu C^\mu  \big) \bigg],\;\;
 \label{58}
\end{eqnarray}
which demonstrate that the action integrals corresponding to ${\cal L}_{(B,{\cal B})}$ 
and ${\cal L}_{(\bar B, \bar {\cal B})}$:  $S_1 = \int d^4 x\,{\cal L}_{(B, {\cal B})}$ 
and $S_2 = \int d^4 x\,{\cal L}_{(\bar B, {\cal B})}$
remain invariant under {\it both} the co-BRST as well as anti-co-BRST symmetry transformations.

Exploiting the theoretical strength of Noether's theorem, we know that the above continuous (anti-)co-BRST symmetry 
transformations lead to the derivation of Noether's conserved currents as:   
\begin{eqnarray}
J^\mu_{ad} &=& \varepsilon^{\mu\nu\eta\kappa}\,\big(m\,\phi_\nu + \bar B_\nu \big) \big(\partial_\eta C_\kappa \big)
+ \Big(\frac{m}{2}\, \varepsilon^{\mu\nu\eta\kappa}\, B_{\eta\kappa} + \tilde \Phi^{\mu\nu} \Big) \big(\partial_\nu C - m C_\nu \big) \nonumber\\
&+& \bar {\cal B} \big(\partial^\mu C - m C^\mu \big) - m \beta \big(\partial^\mu \bar C - m \bar C^\mu \big) 
+ \big(\partial^\mu \bar C^\nu - \partial^\nu \bar C^\mu \big) (\partial_\nu \beta) \nonumber\\
&-& \bar {\cal B}_\nu \big(\partial^\mu C^\nu - \partial^\nu C^\mu \big) - \frac{1}{2}\, (\partial^\mu \beta)\, \rho 
- \frac{1}{2}\, \bar {\cal B}^\mu\, \lambda.
\label{59}
\end{eqnarray}
\begin{eqnarray}
J^\mu_d &=& \varepsilon^{\mu\nu\eta\kappa}\,\big(m\,  \phi_\nu -  B_\nu \big) \big(\partial_\eta \bar C_\kappa \big)
+ \Big(\frac{m}{2}\,\varepsilon^{\mu\nu\eta\kappa}\, B_{\eta\kappa} + \tilde \Phi^{\mu\nu} \Big) \big(\partial_\nu \bar C - m \bar C_\nu \big) \nonumber\\
&-& {\cal B} \big(\partial^\mu \bar C - m \bar C^\mu \big) - m \bar \beta \big(\partial^\mu C - m  C^\mu \big) 
+ \big(\partial^\mu C^\nu - \partial^\nu C^\mu \big) (\partial_\nu \bar \beta)\nonumber\\
&+&  {\cal B}_\nu \big(\partial^\mu \bar C^\nu - \partial^\nu \bar C^\mu \big) + \frac{1}{2}\, (\partial^\mu \bar \beta)\, \lambda
- \frac{1}{2}\, {\cal B}^\mu\, \rho,
\label{60}
\end{eqnarray}
Using the EL-EOMs (quoted in Eqs. (\ref{38}) and (\ref{39})), we can verify that $\partial_\mu J^\mu_{(a)d} = 0$ which demonstrates
the validity of conservation of currents. The above conserved (anti-)co-BRST Noether's currents lead to the definition of the conserved (anti-)co-BRST
charges which are the generators for the continuous (anti-)co-BRST symmetry transformations. These statements can be captured
 in the language of mathematical expressions.  First of all, we note that the (anti-)co-BRST charges $(Q_{(a)d} = \int d^3x\, J^0_{(a)d})$ are 
explicitly expressed as    
\begin{eqnarray}
Q_{ad} &=& \int d^3x \Big[\epsilon^{ijk}\,\big(m\,\phi_i + \bar B_i \big) \big(\partial_j C_k \big)
+ \Big(\frac{m}{2}\, \epsilon^{ijk}\, B_{jk} + \tilde \Phi^{0i} \Big) \big(\partial_i C - m C_i \big) \nonumber\\
&+& \bar {\cal B} \big(\partial^0 C - m C^0 \big) - m \beta \big(\partial^0 \bar C - m \bar C^0 \big) 
+ (\partial_i \beta) \big(\partial^0 \bar C^i - \partial^i \bar C^0 \big) \nonumber\\
&-& \bar {\cal B}_i \big(\partial^0 C^i - \partial^i C^0 \big) - \frac{1}{2}\, (\partial^0 \beta)\, \rho 
- \frac{1}{2}\, \bar {\cal B}^0\, \lambda \Big],
\label{61}
\end{eqnarray}
\begin{eqnarray}Q_d &=& \int d^3x \Big[\epsilon^{ijk}\,\big(m\,  \phi_i -  B_i \big) \big(\partial_j \bar C_k \big)
+ \Big(\frac{m}{2}\,\epsilon^{ijk}\, B_{jk} + \tilde \Phi^{0i} \Big) \big(\partial_i \bar C - m \bar C_i \big) \nonumber\\
&-& {\cal B} \big(\partial^0 \bar C - m \bar C^0 \big) - m \bar \beta \big(\partial^0 C - m  C^0 \big) 
+ (\partial_i \bar \beta) \big(\partial^0 C^i - \partial^\nu C^0 \big) \nonumber\\
&+&  {\cal B}_i \big(\partial^0 \bar C^i - \partial^i \bar C^0 \big) + \frac{1}{2}\, (\partial^0 \bar \beta)\, \lambda
- \frac{1}{2}\, {\cal B}^0\, \rho \Big],
\label{62}
\end{eqnarray}
which are {\it nilpotent} of order {\it two} (i.e. $Q^2_{(a)d} = 0$) as can be explicitly checked by the following relationships
\begin{eqnarray}
&& s_d Q_d = - i\, \big\{Q_d,\, Q_d\big\} = 0, \nonumber\\
&& s_{ad} Q_{ad} = - i\, \big\{Q_{ad},\, Q_{ad} \big\} = 0,
\label{63}
\end{eqnarray}
where the conserved charges $Q_{(a)d}$ have been used as the generators for the (anti-)co-BRST symmetry transformations. 
To be precise, these charges are the generators for any kind of fields (i.e. bosonic/fermionic) as quoted in Eq. (\ref{42}) where
we have to replace $r = a, ab$ by $r = d, ad$ and rest of the symbols denote  their standard meaning(s) as explained earlier.

We end this section with the following crucial remarks. First of all, we observe that the (anti-)co-BRST 
symmetry transformations are supersymmetric-type because they change the bosonic fields into fermionic fields 
and  {\it vice-versa}. Second, the ghost number of a field decreases by {\it one} when we apply the co-BRST symmetry 
transformation on it. On the contrary, the ghost number {\it increases} by one when we apply the anti-co-BRST 
symmetry transformation on the {\it same} field. Finally, the decisive feature of the (anti-)co-BRST symmetry transformations 
is the observation that the {\it total} gauge-fixing term  (for $B_{\mu\nu}$ field) of the theory remains 
{\it invariant} under these transformations.


\section{Bosonic symmetry transformations} \label{Sect. 5}

We have already observed, in our previous {\it two} sections, that there are {\it four } fermionic (i.e. nilpotent) 
symmetry transformations in our present theory. These are the (anti-)BRST and (anti-)co-BRST symmetry transformations which 
are {\it nilpotent} of order two and {\it absolutely anticommuting} in nature. We have also made passing remarks that these fermionic symmetries are 
connected with the exterior and co-exterior derivatives of differential geometry. Thus, it is but natural to think about the existence of 
{\it bosonic} symmetries in our theory. It turns out that $s_\omega = \{s_b,\, s_d\}$ and $s_{\bar \omega} = \{s_{ab},\, s_{ad}\}$  
are the well-defined bosonic symmetry transformations in our theory which can be written as:
\begin{eqnarray}
&& s_\omega B_{\mu\nu} = - \big(\partial_\mu {\cal B}_\nu - \partial_\nu {\cal B}_\mu \big)
 - \varepsilon_{\mu\nu\eta\kappa}\,\partial^\eta B^\kappa, \quad s_\omega C_\mu = \partial_\mu \lambda, 
\quad s_\omega \bar C_\mu = \partial_\mu \rho, \nonumber\\
&& s_\omega \phi_\mu = \partial_\mu {\cal B} - m {\cal B}_\mu, \quad s_\omega \tilde \phi_\mu = \partial_\mu B - m B_\mu, 
\quad s_\omega C = m \lambda, \quad s_\omega \bar C = m \rho,\;\; \nonumber\\
&& s_\omega \big[B_\mu, \bar B_\mu, {\cal B}_\mu, \bar {\cal B}_\mu, B, \bar B, {\cal B}, \bar {\cal B}, \varphi, \tilde \varphi, \beta, \bar \beta,
\rho, \lambda \big] = 0, 
\label{64}
\end{eqnarray}
\begin{eqnarray}
&& s_{\bar \omega} B_{\mu\nu} = - \big(\partial_\mu \bar {\cal B}_\nu - \partial_\nu \bar {\cal B}_\mu \big) 
- \varepsilon_{\mu\nu\eta\kappa}\,\partial^\eta\, \bar B^\kappa, \quad s_{\bar \omega} C_\mu = - \partial_\mu \lambda, 
\;\; s_{\bar \omega} \bar C_\mu = - \partial_\mu \rho, \nonumber\\
&& s_{\bar \omega} \phi_\mu = \partial_\mu \bar {\cal B} - m \bar {\cal B}_\mu, 
\quad s_{\bar \omega} \tilde \phi_\mu = \partial_\mu \bar B - m \,\bar B_\mu, \;\;
s_{\bar \omega} C = - m \lambda, \;\; s_{\bar \omega} \bar C =  - m \rho, \;\; \nonumber\\
&& s_{\bar \omega} \big[B_\mu, \bar B_\mu, {\cal B}_\mu, \bar {\cal B}_\mu, B, \bar B, {\cal B}, 
\bar {\cal B}, \varphi, \tilde \varphi, \beta, \bar \beta, \rho, \lambda \big] = 0. 
\label{65}
\end{eqnarray}
A close look at the above transformations demonstrate that $s_\omega + s_{\bar \omega} = 0$ in {\it all} the transformations 
corresponding to {\it all} the fields of our theory {\it except} in the cases of fields $B_{\mu\nu}$, $\phi_\mu$, and $\tilde \phi_\mu$.
However, it turns out that if we exploit the theoretical potential and validity of the CF-type restrictions (cf. Eq. (\ref{20})), it can be 
readily checked that:
\begin{eqnarray}
&& \big(s_\omega + s_{\bar \omega} \big)\,B_{\mu\nu} = 0, \qquad  \big(s_\omega + s_{\bar \omega} \big) \,\phi_\mu = 0, 
\qquad \big(s_\omega + s_{\bar \omega} \big)\, \tilde \phi_\mu = 0.  
\label{66}
\end{eqnarray}  
Hence, it is clear that, on the submanifold of field space where CF-type restrictions are {\it true}, we have the validity of 
$s_\omega + s_{\bar \omega} = 0$. In other words, there is a {\it unique} bosonic symmetry in our theory 
where a submanifold in the space of fields is {\it defined} by the field equations corresponding 
to the CF-type restrictions (cf. Eq. (\ref{20})).

To verify the above statements, we note that the Lagrangian densities ${\cal L}_{(B, {\cal B} )}$ and  ${\cal L}_{(\bar B, \bar {\cal B} )}$ 
transform, under $s_\omega$ and $s_{\bar \omega}$, as follows
\begin{eqnarray}
s_\omega {\cal L}_{(B, {\cal B})} &=&  - \partial_\mu \, \Big[m\, \varepsilon^{\mu\nu\eta\kappa}
\phi_\nu \big(\partial_\eta B_\kappa \big) + m\, \varepsilon^{\mu\nu\eta\kappa}
\tilde \varphi_\nu \big(\partial_\eta {\cal B}_\kappa \big) - {\cal B}_\nu \big(\partial^\mu { B}^\nu - \partial^\nu B^\mu \big) \nonumber\\
&+& B_\nu \big(\partial^\mu { \cal B}^\nu - \partial^\nu {\cal B}^\mu \big) 
+ {\cal B} \big(\partial^\mu B - m B^\mu \big)  - B \big(\partial^\mu {\cal B} - m {\cal B}^\mu \big)  \nonumber\\
&+& \frac{1}{2}\, \big(\partial^\mu\rho \big)\,\lambda + \frac{1}{2}\, \big(\partial^\mu\lambda \big)\,\rho \Big], 
\label{67}
\end{eqnarray}
\begin{eqnarray}
s_{\bar\omega} {\cal L}_{(\bar B, \bar{\cal B})} &=& - \partial_\mu \, \Big[m\, \varepsilon^{\mu\nu\eta\kappa}
\phi_\nu \big(\partial_\eta \bar B_\kappa \big) + m\, \varepsilon^{\mu\nu\eta\kappa}
\tilde\varphi_\nu\big(\partial_\eta \bar{\cal B}_\kappa \big) + \bar{\cal B}_\nu \big(\partial^\mu {\bar B}^\nu-\partial^\nu \bar B^\mu \big) \nonumber\\
&-& {\bar B}_\nu \big(\partial^\mu \bar{ \cal B}^\nu - \partial^\nu \bar{\cal B}^\mu \big) 
- \bar{\cal B} \big(\partial^\mu \bar B - m \bar B^\mu \big)  +  {\bar B} \big(\partial^\mu \bar {\cal B} - m \bar{\cal B}^\mu \big) \nonumber\\ 
&-& \frac{1}{2}\, \big(\partial^\mu\rho \big)\,\lambda - \frac{1}{2}\, \big(\partial^\mu\lambda \big)\,\rho \Big], 
\label{68}
\end{eqnarray}
which demonstrate that the action integrals corresponding to ${\cal L}_{(B, {\cal B} )}$ and ${\cal L}_{(\bar B, \bar {\cal B} )}$:
 $S_1 = \int d^4x\, {\cal L}_{(B, {\cal B} )}$ and  
$S_2 = \int d^4x\,{\cal L}_{(\bar B, \bar {\cal B} )}$ remain invariant under $s_\omega$ and $s_{\bar\omega}$, respectively. 
In other words, we have the bosonic symmetries $s_\omega$ and $s_{\bar\omega}$ for the {\it coupled} Lagrangian densities 
${\cal L}_{(B, {\cal B} )}$ and  ${\cal L}_{(\bar B, \bar {\cal B} )}$ (cf. Eqs. (\ref{67}), (\ref{68})), respectively. 
However, these bosonic symmetries are the {\it symmetry} transformations of {\it both} the Lagrangian densities on the 
submanifold in the field space defined by the CF-type restrictions as can be seen by the following explicit transformations 
(i.e. from the expressions for $s_{\bar \omega} {\cal L}_{(B, {\cal B} )}$ and $s_\omega {\cal L}_{(\bar B, \bar {\cal B} )}$), namely;
\begin{eqnarray}
s_{\bar\omega} {\cal L}_{( B, {\cal B})} &=& - \partial_\mu \, \Big[m\, \varepsilon^{\mu\nu\eta\kappa}
\phi_\nu \big(\partial_\eta \bar B_\kappa\big) + m\, \varepsilon^{\mu\nu\eta\kappa}
\tilde\varphi_\nu \big(\partial_\eta \bar{\cal B}_\kappa \big) - {\cal B}_\nu \big(\partial^\mu {\bar B}^\nu-\partial^\nu \bar B^\mu \big) \nonumber\\ 
&+& B_\nu \big(\partial^\mu \bar{ \cal B}^\nu - \partial^\nu \bar {\cal B}^\mu \big)
+ {\cal B} \big(\partial^\mu \bar B - m \bar B^\mu \big)  - B \big(\partial^\mu \bar{\cal B} - m \bar{\cal B}^\mu \big) \nonumber\\ 
&-& \frac{1}{2}\, \big(\partial^\mu\rho\big)\,\lambda - \frac{1}{2}\, \big(\partial^\mu\lambda \big)\,\rho \Big] 
+ \partial_\mu \big[B_\nu + \bar B_\nu + \partial_\nu \varphi \big] \, 
\big(\partial^\mu \bar{\cal B}^\nu  - \partial^\nu \bar{\cal B}^\mu \big) \nonumber\\ 
&-& \partial_\mu \big[{\cal B}_\nu + \bar{\cal B}_\nu  + \partial_\nu \tilde\varphi \big] \, \big(\partial^\mu \bar{ B}^\nu
 - \partial^\nu \bar{ B}^\mu \big) \nonumber\\ 
 &-& \big[ \partial_\mu \big(B + \bar B + m\,\varphi\big) - m\, \big(B_\mu + \bar B_\mu  + \partial_\mu \varphi\big) \big] \, 
\big(\partial^\mu \bar{\cal B} - m \bar{\cal B}^\mu \big) \nonumber\\ 
&+&  \big[ \partial_\mu \big({\cal B} + \bar{\cal B} + m\,\tilde\varphi \big) - m\, \big({\cal B}_\mu + \bar {\cal B}_\mu 
 + \partial_\mu \tilde\varphi \big)\big] \,\big(\partial^\mu \bar B - m \bar B^\mu \big),  
\label{69} 
\end{eqnarray}
\begin{eqnarray}
s_{\omega} {\cal L}_{(\bar B, \bar {\cal B})} &=&  - \partial_\mu \, \Big[m\, \varepsilon^{\mu\nu\eta\kappa}
\phi_\nu \big(\partial_\eta B_\kappa \big) +  m\, \varepsilon^{\mu\nu\eta\kappa}
\tilde\varphi_\nu \big(\partial_\eta {\cal B}_\kappa \big) 
+ \bar {\cal B}_\nu \big(\partial^\mu  B^\nu - \partial^\nu B^\mu \big) \nonumber\\
&-& \bar B_\nu \big(\partial^\mu { \cal B}^\nu - \partial^\nu {\cal B}^\mu \big) 
- \bar {\cal B} \big(\partial^\mu B - m B^\mu \big)  +  {\bar B} \big(\partial^\mu {\cal B} - m {\cal B}^\mu \big) \nonumber\\ 
&+& \frac{1}{2}\, \big(\partial^\mu\rho \big)\,\lambda + \frac{1}{2}\, \big(\partial^\mu\lambda \big)\,\rho \Big] 
- \partial_\mu \big[B_\nu + \bar B_\nu + \partial_\nu \varphi \big] \, \big(\partial^\mu {\cal B}^\nu
 - \partial^\nu {\cal B}^\mu \big) \nonumber\\ 
&+& \partial_\mu \big[{\cal B}_\nu + \bar{\cal B}_\nu  + \partial_\nu \tilde\varphi \big] \, 
\big(\partial^\mu B^\nu  - \partial^\nu B^\mu\big) \nonumber\\ 
 &+& \big[\partial_\mu \big(B + \bar B + m\,\varphi \big) - m\, \big(B_\mu + \bar B_\mu  + \partial_\mu \varphi \big)\big] \, 
\big(\partial^\mu {\cal B} - m {\cal B}^\mu \big) \nonumber\\ 
&-&  \big[ \partial_\mu \big({\cal B} + \bar{\cal B} + m\,\tilde\varphi \big) - m\, \big({\cal B}_\mu + \bar {\cal B}_\mu 
 + \partial_\mu \tilde\varphi \big)\big] \, \big(\partial^\mu B - m  B^\mu \big), 
\label{70}
\end{eqnarray}
which prove that {\it both} the Lagrangian densities respect {\it both} the symmetry transformations $s_\omega$ and $s_{\bar \omega}$.
However, the operator relationship $s_\omega + s_{\bar \omega} = 0$ on the on the submanifold in the field space 
(where the CF-type restrictions are true) implies that  we have a {\it unique} bosonic symmetry transformation 
(i.e. either $s_\omega$ or $s_{\bar \omega}$) out of the two.

According to Noether's theorem, the above continuous bosonic symmetry transformations lead to the 
following expressions for the conserved currents:
\begin{eqnarray}
J^\mu_\omega &=& \varepsilon^{\mu\nu\eta\kappa}\, \big(m\,\phi_\nu - B_\nu \big) \big(\partial_\eta B_\kappa \big) 
+  \varepsilon^{\mu\nu\eta\kappa}\,\big(m\,\tilde\phi_\nu - {\cal B}_\nu \big) \big(\partial_\eta {\cal B}_\kappa\big) \nonumber\\
&-& \big(m {\cal B}_\nu - \partial_\nu {\cal B} \big)\, \big(m B^{\mu\nu} - \Phi^{\mu\nu} \big) 
- \big(m B_\nu - \partial_\nu B \big)\, \Big(\frac{ m}{2}\, \varepsilon^{\mu\nu\eta\kappa}\, B_{\eta\kappa} + {\tilde\Phi}^{\mu\nu}\Big) \nonumber\\
&-& \big(\partial^\mu \bar C^\nu - \partial^\nu \bar C^\mu\big)\big(\partial_\nu \lambda\big) 
+ \big(\partial^\mu  C^\nu - \partial^\nu  C^\mu\big) \big(\partial_\nu \rho \big) \nonumber\\
&+& m \big(\partial^\mu \bar C - m\,\bar C^\mu\big) \lambda - m \big(\partial^\mu C - m \,  C^\mu\big) \rho, 
\label{71}
\end{eqnarray}
\begin{eqnarray}
J^\mu_{\bar\omega}&=& \varepsilon^{\mu\nu\eta\kappa}\, \big(m\,\phi_\nu + \bar B_\nu \big) \, \big(\partial_\eta \bar B_\kappa \big) 
+ \varepsilon^{\mu\nu\eta\kappa}\,\big(m\, \tilde\phi_\nu + \bar{\cal B}_\nu \big) \, \big(\partial_\eta \bar{\cal B}_\kappa \big) \nonumber\\
&-& \big(m \bar{\cal B}_\nu - \partial_\nu \bar{\cal B} \big)\, \big( m B^{\mu\nu} - \Phi^{\mu\nu} \big) 
- \big(m {\bar B}_\nu - \partial_\nu \bar B \big)\,
 \Big(\frac{ m}{2}\, \varepsilon^{\mu\nu\eta\kappa}\, B_{\eta\kappa} + {\tilde \Phi}^{\mu\nu} \Big) \nonumber\\
&+& \big(\partial^\mu \bar C^\nu - \partial^\nu \bar C^\mu\big) \big(\partial_\nu \lambda \big) 
- \big(\partial^\mu  C^\nu - \partial^\nu  C^\mu\big) \big(\partial_\nu \rho \big)\nonumber\\ 
&-& m \big(\partial^\mu \bar C - m\,\bar C^\mu\big) \lambda + m \big(\partial^\mu C - m \,  C^\mu\big) \rho. 
\label{72}
\end{eqnarray}
It is very important to point out that the above currents are {\it not} independent of each other (on the 
on the submanifold in the field space where the CF-type restrictions are satisfied 
(cf. Eq. (\ref{20})). This is due to the fact that we have the following exact relationship: 
\begin{eqnarray}
J^\mu_\omega +  J^\mu_{\bar \omega} = 0.
\label{73}
\end{eqnarray}
As a consequence of the above observation, we note that the following charges $Q_\omega = \int d^3x\, J^0_\omega$
and $Q_{\bar \omega} = \int d^3x\, J^0_{\bar \omega}$, namely;
\begin{eqnarray}
Q_\omega &=& \int d^3x \Big[\epsilon^{ijk}\, \big(m\,\phi_i - B_i \big) \big(\partial_j B_k \big) 
+ \epsilon^{ijk}\,\big(m\,\tilde\phi_i - {\cal B}_i \big) \big(\partial_j {\cal B}_k\big) \nonumber\\
&-& \big(m {\cal B}_i - \partial_i {\cal B} \big)\, \big(m B^{0i} - \Phi^{0i} \big) 
- \big(m B_i - \partial_i B \big)\, \Big(\frac{ m}{2}\, \epsilon^{ijk}\, B_{jk} + {\tilde\Phi}^{0i}\Big) \nonumber\\
&-& \big(\partial^0 \bar C^i - \partial^i \bar C^0\big)\big(\partial_i \lambda\big) 
+ \big(\partial^0  C^i - \partial^i  C^0\big) \big(\partial_i \rho \big) \nonumber\\
&+& m \big(\partial^0 \bar C - m\,\bar C^0\big) \lambda - m \big(\partial^0 C - m \,  C^0\big) \rho \Big],
\label{74}
\end{eqnarray}
\begin{eqnarray}
Q_{\bar\omega}&=& \int d^3x \Big[\epsilon^{ijk}\, \big(m\,\phi_i + \bar B_i \big) \, \big(\partial_j \bar B_k \big) 
+ \epsilon^{ijk}\,\big(m\, \tilde\phi_i + \bar{\cal B}_i \big) \, \big(\partial_j \bar{\cal B}_k \big) \nonumber\\
&-& \big(m \bar{\cal B}_i - \partial_i \bar{\cal B} \big)\, \big( m B^{0i} - \Phi^{0i} \big) 
- \big(m {\bar B}_i - \partial_i \bar B \big)\,
 \Big(\frac{ m}{2}\, \epsilon^{ijk}\, B_{jk} {\tilde\Phi}^{0i} \Big) \nonumber\\
&+& \big(\partial^0 \bar C^i - \partial^i \bar C^0 \big) \big(\partial_i \lambda \big) 
- \big(\partial^0  C^i - \partial^i  C^0 \big) \big(\partial_i \rho \big)\nonumber\\ 
&-& m \big(\partial^0 \bar C - m\,\bar C^0 \big) \lambda + m \big(\partial^0 C - m \,  C^0 \big) \rho \Big],
\label{75}
\end{eqnarray}
are also  {\it not} independent of each other if we exploit the beauty and strength of the CF-type restrictions (cf. Eq. (\ref{20})). 
In fact, we lay emphasis on the fact that we have a {\it unique} bosonic charge 
$Q_\omega = \{Q_b, \, Q_d \} = -  \{Q_{ab}, \, Q_{ad} \} = - \,Q_{\bar \omega}$ on the on the 
submanifold in the  space of fields
where the CF-type restrictions (cf. Eq. (\ref{20})) are {\it true}. We point out that the currents (\ref{71}) and (\ref{72}) are conserved 
(i.e. $\partial_\mu J^\mu_\omega = \partial_\mu J^\mu_{\bar \omega} = 0$) due to the EL-EOMs that are  given in Eqs. (\ref{38}) and (\ref{39}) 
(cf. Section \ref{Sect. 3} for more details).


\section{Ghost-scale symmetry and discrete symmetries} \label{Sect. 6}

In addition to the {\it five} symmetries (i.e. {\it four} fermionic and one {\it unique  bosonic} symmetries), we have a {\it continuous}
symmetry in our theory which is known as the ghost-scale symmetry transformation. This symmetry is confined to the fields present in the 
Faddeev--Popov ghost sector of the Lagrangian densities ${\cal L}_{(B, {\cal B})}$ and ${\cal L}_{(\bar B, \bar {\cal B})}$. The 
characteristic feature of the ghost-scale symmetry  transformation is the fact that {\it only} the (anti-)ghost fields 
transform (according to their ghost numbers) and the {\it rest} of the {\it ordinary} fields of the theory do {\it not} transform at all. For our theory, 
we have the following ghost-scale symmetry transformations (with $\Omega = $ spacetime independent scale parameter), namely;  
\begin{eqnarray}
&& C_\mu \to e^{+ \Omega} \, C_\mu, \quad \bar C_\mu \to e^{- \Omega} \, \bar C_\mu, \quad
 C \to e^{+ \Omega} \, C, \quad  \bar C \to e^{- \Omega} \, \bar C, \nonumber\\
&& \beta \to e^{+2 \Omega} \, \beta, \quad \bar \beta \to e^{-2 \Omega} \, \bar \beta, \quad \lambda \to e^{+ \Omega} \, \lambda, 
\quad \rho \to e^{-  \Omega} \, \rho,\quad \Psi \to e^0\, \Psi, \;\;\;\;
\label{76}
\end{eqnarray}
where $\Psi (= B_{\mu\nu}, \phi_\mu, \tilde\phi_\mu, B_\mu, \bar B_\mu,  {\cal B}_\mu, \bar{\cal B}_\mu,  
B, \bar B, {\cal B}, \bar{\cal B},  \varphi, \tilde\varphi)$ is the {\it generic ordinary field} of the theory with ghost number equal to 
zero. In the above, the {\it numerals} in the exponent correspond to the {\it ghost numbers} for the specific (anti-)ghost field under consideration.
The infinitesimal version ($s_g$) of the above ghost-scale symmetry transformations (cf. Eq. (\ref{76})) is 
\begin{eqnarray}
&& s_g C_\mu = + \, C_\mu, \qquad  s_g \bar C_\mu = - \, \bar C_\mu, \qquad s_g C = + \, C, \qquad s_g \bar C = - \, \bar C, \nonumber\\
&&  s_g \beta = + 2 \, \beta, \qquad  s_g \bar \beta = - 2 \, \bar \beta,  \qquad  s_g \rho = - \, \rho,\qquad s_g \lambda = + \, \lambda,\nonumber\\
&& s_g \big[ B_{\mu\nu}, \phi_\mu, \tilde\phi_\mu, B_\mu, \bar B_\mu, {\cal B}_\mu, \bar{\cal B}_\mu,   B, \bar B, 
{\cal B}, \bar{\cal B}, \varphi, \tilde\varphi \big] = 0, \nonumber\\
\label{77}
\end{eqnarray}
where, for the sake of brevity, we have taken the constant (i.e. spacetime independent) global scale parameter $\Omega =1$. It is elementary to check that 
$s_g {\cal L}_{(B, {\cal B})} = s_g {\cal L}_{(\bar B, \bar{\cal B})} = 0$ which demonstrate that the coupled Lagrangian densities (as well 
as their corresponding action integrals) remain invariant under the infinitesimal version ($s_g$) of the ghost-scale transformations 
(cf. Eqs. (\ref{76}), (\ref{77})) which are continuous symmetry transformations.

We exploit now the Noether theorem  to derive the expressions for the conserved current and charge for 
the infinitesimal version of the ghost-scale symmetry transformations as:
\begin{eqnarray}
J^\mu_g &=& - \big(\partial^\mu \bar C^\nu - \partial^\nu \bar C^\mu \big) C_\nu
- \big(\partial^\mu C^\nu - \partial^\nu C^\mu \big) \bar C_\nu + \big(\partial^\mu \bar C - m\, \bar C^\mu\big) C \nonumber\\
&+&  \big(\partial^\mu C - m\, C^\mu\big) C -   \beta (\partial^\mu \bar \beta) 
+ \bar \beta (\partial^\mu \beta) - \frac{1}{2}\, C^\mu\, \rho + \frac{1}{2}\, \bar C^\mu\, \lambda,
\label{78}
\end{eqnarray}
\begin{eqnarray}
Q_g &=& \int d^3x\, J^0_g \nonumber\\
&=& - \int d^3x \Big[\big(\partial^0 \bar C^i - \partial^i \bar C^0 \big) C_i
+ \big(\partial^0 C^i - \partial^i C^0 \big) \bar C_i - \big(\partial^0 \bar C - m\, \bar C^0\big) C \nonumber\\ 
&-&  \big(\partial^0 C - m\, C^0 \big) C +  \beta (\partial^0 \bar \beta) 
- \bar \beta (\partial^0 \beta) + \frac{1}{2}\, C^0\, \rho - \frac{1}{2}\, \bar C^0 \, \lambda \Big].
\label{79}
\end{eqnarray}
It is quite straightforward to note that the conservation of the current and charge is hidden in the proof $\partial_\mu J^\mu_g = 0$.
For the proof of the latter (i.e. $\partial_\mu J^\mu_g = 0$), we have to use the EL-EOMs that have been listed in Eqs. (\ref{38}) and (\ref{39}).
We also note that the charge $Q_g$ is the generator for the infinitesimal transformation $(s_g)$  when we use the general expression 
(cf. Eq. (\ref{42})) for the relationship between the continuous symmetry transformation $s_r$ and the generator $Q_r$. 
For the case of ghost-scale infinitesimal symmetry transformation, it is clear that we have to take $r =g$ in the general expression (cf. Eq. (\ref{42})).

We end our discussion on the ghost-scale infinitesimal  symmetry transformations with the following remarks. First of all, 
we note that the following are true, namely;
\begin{eqnarray}
&& s_g Q_b = - i \, \big[Q_b, \, Q_g \big] = + \,Q_b, \qquad s_g Q_d = - i \, \big[Q_d, \, Q_g \big] = - \, Q_d, \nonumber\\
&& s_g Q_{ab} = - i \, \big[Q_{ab}, \, Q_g \big] = -\, Q_{ab}, \qquad s_g Q_{ad} = - i \, \big[Q_{ad}, \, Q_g \big] = +\, Q_{ad}, \nonumber\\
&& s_g Q_\omega = - i \, \big[Q_g, \, Q_\omega \big] = 0, \qquad s_g Q_g = - i \, \big[Q_g, \, Q_g \big] = 0.
\label{80}
\end{eqnarray}
In the above, we have utilized the key concepts of symmetry principle which provides a connection between the continuous symmetry transformation
and corresponding conserved charge as its generator. The above algebra is very important because if we define the ghost 
number of a state $|\psi \rangle_n$ in the quantum Hilbert space of states by $n$ (i.e. $i\,Q_g |\psi \rangle_n = n\, |\psi \rangle_n$) 
which is nothing but the eigenvalue of the operator ``$i\,Q_g$", we observe the 
following:
\begin{eqnarray}
&& i\, Q_g\, Q_b\, |\psi \rangle_n = (n + 1)\, Q_b\, |\psi \rangle_n, 
\qquad i\, Q_g\, Q_{ad}\, |\psi \rangle_n = (n + 1)\, Q_{ad}\, |\psi \rangle_n, \nonumber\\
&& i\, Q_g\, Q_{ab} \,|\psi \rangle_n = (n - 1)\, Q_{ab}\, |\psi \rangle_n, 
\qquad i\, Q_g\, Q_d\, |\psi \rangle_n = (n - 1)\, Q_d\, |\psi \rangle_n, \nonumber\\
&& i\, Q_g\, Q_\omega\, |\psi \rangle_n = n \,Q_\omega \,|\psi \rangle_n.
\label{81}
\end{eqnarray}
The above relations demonstrate that the ghost numbers of states $(Q_b |\psi \rangle_n$, $Q_d |\psi \rangle_n$, $Q_\omega |\psi \rangle_n$) 
and  $(Q_{ad} |\psi \rangle_n$, $Q_b |\psi \rangle_n$, $Q_\omega |\psi \rangle_n$) are $(n + 1)$, $(n - 1)$
and $n$, respectively. This observation would play very important role in our Section \ref{Sect. 7}. Second, we observe that 
the ghost charge is {\it bosonic} in nature despite the fact that, in our theory, there are fermionic as well 
as bosonic ghost fields (that are primarily needed for the validity of unitarity at the quantum level).

Now we dwell a bit on the generalization of the discrete symmetry transformations (\ref{22}) that are present at the gauge-fixed 
Lagrangian  densities (\ref{17}) and (\ref{18}). We note that the transformations (\ref{22}) are {\it amongst} the bosonic fields 
of our theory. As far as the (anti-)BRST and (anti-)co-BRST Lagrangian densities ${\cal L}_{(B, {\cal B})}$  and 
${\cal L}_{(\bar B, \bar {\cal B})}$ are concerned , we observe that under the following {\it discrete} symmetry transformations
\begin{eqnarray}
&& \phi_\mu \to \pm\, i\, \tilde \phi_\mu, \; \qquad \tilde \phi_\mu \to \mp\, i\, \phi_\mu, 
\qquad \varphi \to \pm \,i\, \tilde \varphi, \;\;\qquad \tilde\varphi \to \mp\, i\, \varphi, \nonumber\\
&& B_\mu \to \pm \,i\, {\cal B}_\mu, \qquad {\cal B}_\mu \to \mp\, i\, B_\mu, 
\qquad B \to \pm\, i \,{\cal B}, \qquad {\cal B} \to \mp\, i\, B, \nonumber\\
&& \bar B_\mu \to \pm \,i\, \bar {\cal B}_\mu, \qquad \bar {\cal B}_\mu \to \mp \,i\,\bar  B_\mu, 
\qquad \bar B \to \pm\, i\, \bar {\cal B}, \qquad \bar {\cal B} \to \mp\, i\, \bar B, \nonumber\\
&& C_\mu \to \pm\, i\, \bar C_\mu, \qquad  \bar C_\mu \to \pm\, i\, C_\mu, \qquad C \to \pm \,i\, \bar C, \qquad \bar C \to \pm\, i\, C, \nonumber\\
&& \rho \to \mp\, i\,  \lambda, \;\quad\qquad  \lambda \to \mp\, i\, \rho, \; \quad\qquad \beta \to \pm \,i\, \bar \beta, 
\quad\quad \bar \beta \to \mp \,i\, \beta, \nonumber\\
&& B_{\mu\nu}\to \mp\, \frac{i}{2}\, \varepsilon_{\mu\nu\eta\kappa} B^{\eta\kappa}, \qquad B_{\mu\nu} B^{\mu\nu} \to B_{\mu\nu} B^{\mu\nu},
\label{82}
\end{eqnarray}
the above Lagrangian densities {\it remain} invariant. A close look at the above discrete symmetry transformations 
demonstrates that actually there are  {\it two} discrete symmetry transformations that are hidden in {\it it} depending
on the upper and lower signatures that are associated with the transformations of the  fields. It is also clear that the kinetic and 
gauge-fixing parts of the coupled Lagrangian densities have a separate set of discrete symmetry transformations (cf. Eq. (\ref{22}))
{\it than} the ghost part of the (anti-)BRST and (anti-)co-BRST invariant Lagrangian densities 
${\cal L}_{(B, {\cal B})}$ and ${\cal L}_{(\bar B, \bar {\cal B})}$ of our 4D theory.

We end our discussion on the {\it discrete} symmetry transformations with the remark that these transformations 
would  play a {\it decisive} role in the next section (i.e. Section \ref{Sect. 7}) where we shall discuss the algebraic structures 
of the {\it operator} forms of the charges and symmetries and establish their connection with the algebra of cohomological operators
of differential geometry.


\section{Algebraic structures: Symmetry transformation operators and  conserved charges as operators} \label{Sect. 7}

It is clear that we have {\it six} continuous symmetries in the theory out of which {\it four} are fermionic and {\it two} are bosonic. 
In addition, we have established and shown the existence of a couple of discrete symmetries in the theory. 
One can check that the continuous symmetry transformations (i.e. $s_{(a)b}$, $s_{(a)d}$, $s_\omega$, $s_g$)
satisfy the following algebra, namely;
\begin{eqnarray}
&& s^2_b = 0, \qquad s^2_{ab} = 0, \qquad s^2_d = 0, \qquad s^2_{ad} = 0, \qquad s_{\omega} = \{s_b, \, s_d\} = - s_{\bar \omega},  \nonumber\\
&& \{s_b, \, s_{ab}\} = 0, \qquad \{s_d, \, s_{ad}\} = 0, \qquad \{s_b, \, s_{ad}\} = 0, \qquad \{s_{ab}, \, s_d\} = 0, \nonumber\\
&& [s_g, \, s_b] = + s_b, \quad [s_g, \, s_{ab}] = - s_{ab}, \quad [s_g, \, s_d] = - s_d, \quad [s_g, \, s_{ad}] = + s_{ad}, \nonumber\\
&& [s_\omega, s_r] = 0, \qquad r = b, {ab}, d, {ad}, g, \qquad  s_{(a)d} = \pm * s_{(a)b} *.
\label{83}
\end{eqnarray}
The above algebra demonstrates that $s_\omega$ is like a Casimir operator (but not in the Lie algebraic sense). However,
the validity of the above algebra requires that the CF-type restrictions (cf. Eq. (\ref{20})) are satisfied. In other words, the above algebra
is satisfied on the submanifold of the field space where the CF-type restrictions 
(\ref{20})  are satisfied. In fact, the CF-type restrictions (cf. Eq. (\ref{20})) are the field equations that {\it fully} define the {\it submanifold}.

One of the crucial relationships that the above symmetry operators satisfy (in their operator form) is 
\begin{eqnarray}
s_{(a)d} = \pm \,*\, s_{(a)b}\, *,
\label{84}
\end{eqnarray}
where $*$ is nothing but the discrete symmetry transformations we have discussed in our previous section (cf. Eq. (\ref{82})).
Thus, we note that it is an {\it interplay} between the underlying discrete as well as continuous symmetries of 
the theory that provide the physical realization of the celebrated relationship between the (co-)exterior 
derivatives (i.e. $\delta = \pm \,* \,d\, *$) of the cohomological operators of differential geometry. We further 
note that the algebra (\ref{83}) provides the physical realization of the Hodge algebra \cite{eug,mukh,van,des} that 
is satisfied by the de Rham cohomological operators of differential geometry, namely;
\begin{eqnarray}
&& d^2 =0 ,\qquad \delta^2 = 0, \qquad \big\{d, \, \delta \big\} = \Delta, \nonumber\\
&& \big[\delta, \, d \big] = 0, \qquad \big[\Delta, \, \delta \big] = 0, \qquad \delta = \pm \,* d *,
\label{85}
\end{eqnarray}
where the (co-)exterior derivatives $(\delta) d$ and the Laplacian operator $\Delta$ constitute a set $(d,\, \delta, \Delta)$
of the cohomological operators of differential geometry \cite{eug,mukh,van,des}.

We have defined and discussed  the conserved currents and charges (in our previous section) which are the {\it generators}
of the continuous symmetry transformations. It turns out that these charges satisfy exactly the {\it same } algebra
as the symmetry operators (cf. Eq. (\ref{83})). In other words, we have the following 
\begin{eqnarray}
&& Q^2_b = 0, \qquad\quad Q^2_{ab} = 0,\quad \qquad Q^2_d = 0, \qquad Q^2_{ad} = 0, \nonumber\\
&& \{Q_b, \, Q_{ab}\} = 0, \quad \{Q_d, \, Q_{ad}\} = 0, \quad \{Q_b, \, Q_{ad}\} = 0, \quad \{Q_{ab}, \, Q_d\} = 0, \nonumber\\
&& i\,[Q_g, \, Q_b] = + \,Q_b, \qquad i[Q_g, \, Q_{ab}] = - Q_{ab}, \qquad i\,[Q_g, \, Q_d] = -\, Q_d, \nonumber\\
&& i\,[Q_g, \, Q_{ad}] = +\, Q_{ad}, \qquad  Q_{\omega} = \{Q_b, \, Q_d\} = - \,\{Q_{ab}, \, Q_{ad}\}, \nonumber\\
&& [Q_\omega, \,Q_r] = 0, \qquad r = b, {ab}, d, {ad}, g,
\label{86}
\end{eqnarray}
which demonstrate that $Q_\omega$ is just like the Casimir operator for the whole algebra ({\it but} not in the Lie algebraic sense).
The above algebra is {\it also} reminiscent of the algebra satisfied by the de Rham cohomological operators of differential geometry
(cf. Eq. (\ref{85})). A close look at (\ref{86}) shows that we have the following two-to-one mapping from the charges to cohomological operators
\begin{eqnarray}
&& \big(Q_b, \, Q_{ad} \big) \longrightarrow d, \qquad \big(Q_d, \, Q_{ab} \big) \longrightarrow \delta, \qquad 
(Q_\omega,\, -\, Q_{\bar \omega})  \longrightarrow \Delta,
\label{87}
\end{eqnarray}
from the {\it physically} well-defined conserved charges corresponding to the {\it continuous} and {\it infinitesimal} 
symmetry transformations to the {\it mathematically} well-defined  de Rham cohomological operators of differential geometry.

As a consequence of the above realizations, one obtains a Hodge decomposition theorem  \cite{eug,mukh,van,des} 
in the quantum Hilbert space of states for any arbitrary state $|\omega \rangle_n$ with the ghost number 
$n$ (i.e. $i\,Q_g\, |\omega\rangle_n = n\, |\omega\rangle_n$)
\begin{eqnarray}
|\omega \rangle_n &=& |h \rangle_n + Q_b \,|\alpha \rangle_{n -1} + Q_d \,|\beta \rangle_{n+1} \nonumber\\
&=& |h \rangle_n + Q_{ad}\, |\alpha \rangle_{n -1} + Q_b\, |\beta \rangle_{n+1},
\label{88}
\end{eqnarray}
where $|h \rangle_n$ is the harmonic state (i.e. $Q_\omega\, |h \rangle_n  = 0 \Rightarrow Q_b\, |h \rangle_n  = 0$, $Q_d\, |h \rangle_n = 0$ and
 $Q_{ab}\, |h \rangle_n  = 0$, $Q_{ad}\, |h \rangle_n = 0$), $ Q_b\, |\alpha \rangle_{n-1}$ is the BRST-exact state and 
 $Q_d\, |\beta \rangle_{n+1}$ is the BRST co-exact state in the quantum Hilbert space of states. The most symmetric state 
(i.e. physical state) is the harmonic state which is annihilated by {\it all}  the conserved charges of the theory 
(i.e. $Q_{(a)b} \,|phys \rangle  = 0$, $Q_{(a)}d \,|phys \rangle = 0$, $Q_\omega\, |phys \rangle = 0$). Here the state 
$|phys \rangle$ is nothing but the harmonic state $|h\rangle_n$ that must be chosen as the {\it physical state} 
(i.e. $|phys \rangle$). At the {\it physical} level, such a state would be annihilated by, at least, BRST charge and 
co-BRST charge which would lead to the annihilation of the physical state by the {\it first-class} constraints.
We have performed such kind of computations in our earlier works \cite{rpm0,rk1,rpm1,rpm2,rpm3,rpm4,rpm5,rpm6}. 
The same kind of analysis can be repeated  for our system under consideration, too.

We wrap up this section with the remark that the symmetry operators and/or the conserved charges of our theory 
provide the physical realizations of the cohomological operators of differential geometry. Hence, our 4D 
{\it massive} Abelian 2-form gauge theory is a tractable field-theoretic example for the Hodge theory (which leads to the 
existence and emergence of the fields/particles with {\it negative} kinetic terms that we discuss below).


\section{Comments on the negative kinetic terms: Physical aspects} \label{Sect. 8}

We have demonstrated that the free 4D {\it massive } Abelian 2-form gauge theory is a tractable field-theoretic model for the Hodge 
theory where the {\it discrete} and {\it continuous} symmetry transformations  play pivotal roles in providing the physical 
realizations of {\it all} the mathematical ingredients connected with the set of well-known de Rham cohomological operators 
of differential geometry at the {\it algebraic} level. A decisive role is played by the {\it discrete} symmetry transformations of 
our theory where we note that the pseudo-scalar and  axial-vector fields are introduced with {\it negative } kinetic terms {\it but} 
with proper definition  of mass.

Let us focus on the explicit expression for the kinetic term of the Abelian 2-form field $(B_{\mu\nu})$ in our present discussion:
This term is as follows: 
\begin{eqnarray}
\frac{1}{2}\,{\cal B}_\mu {\cal B}^\mu - {\cal B}^\mu \bigg(\frac{1}{2}\, \varepsilon_{\mu\nu\eta\kappa}\, \partial^\nu B^{\eta\kappa}
- \frac{1}{2}\, \partial_\mu \tilde \varphi  + m\, \tilde \phi_\mu \bigg).
\label{89}
\end{eqnarray}
Using the EL-EOM, we observe that ${\cal B}_\mu = \frac{1}{2}\, \varepsilon_{\mu\nu\eta\kappa}\, \partial^\nu B^{\eta\kappa}
- \frac{1}{2}\, \partial_\mu \tilde \varphi  + m\, \tilde \phi_\mu$. Thus, this kinetic tem is, primarily, equal to
$ - \frac{1}{2}\, {\cal B}_\mu {\cal B}^\mu$ {\it on-shell}. When we substitute the expression  for  ${\cal B}_\mu$ into it, we 
obtain the following {\it kinetic} terms for the $B_{\mu\nu}$ and pseudo-scalar fields (along with other useful terms)
\begin{eqnarray}
\frac{1}{12}\, H^{\mu\nu\eta} H_{\mu\nu\eta} - \frac{1}{8}\, \partial^\mu \tilde \varphi \,\partial_\mu \tilde \varphi
 - \frac{m}{4}\, \varepsilon^{\mu\nu\eta\kappa}\, B_{\mu\nu} \tilde \Phi_{\eta\kappa} \nonumber\\
 - \frac{m^2}{2}\, \tilde \phi_\mu \,\tilde \phi^\mu
 + \frac{m}{2}\, \tilde \phi^\mu \,\partial_\mu \tilde \varphi,
 \label{90}
\end{eqnarray}
where the total spacetime derivative terms have been dropped due to obvious reasons. The above equation demonstrates that 
we have obtained the {\it correct} signature (with proper numerical factor) for the Kalb-Ramond Lagrangian density of the antisymmetric  tensor gauge field 
$(B_{\mu\nu})$. However, the corresponding signature of the kinetic term for the {\it pseudo-scalar} field is 
{\it negative}. Thus, the pseudo-scalar field turns up in our theory with a negative kinetic term.

The above observation should be contrasted with the gauge-fixing term for the 4D Abelian 2-form gauge field $(B_{\mu\nu})$.
The linearized version of this term is 
\begin{eqnarray}
B^\mu \bigg(\partial^\nu B_{\nu\mu} - \frac{1}{2}\, \partial_\mu \varphi + m \,\phi_\mu \bigg) - \frac{1}{2}\, B^\mu B_\mu,
\label{91}
\end{eqnarray}         
where $B_\mu  = \partial^\nu B_{\nu\mu} - \frac{1}{2}\, \partial_\mu \varphi + m \,\phi_\mu$. The above term is basically equal to 
$\frac{1}{2}\, B^\mu B_\mu$  {\it on-shell}. It is evident that the kinetic term for the pure scalar field $\varphi$ and the 
gauge-fixing term for the $B_{\mu\nu}$ field appear in the theory as
\begin{eqnarray}
\frac{1}{2}\, \Big(\partial^\nu B_{\nu\mu} \Big)\Big(\partial_\rho B^{\rho\mu} \Big)
 + \frac{1}{8}\, \partial^\mu \varphi\, \partial_\mu \varphi + \frac{m^2}{2}\, \phi^\mu \phi_\mu  \nonumber\\
- \frac{m}{2}\, \phi^\mu \,\partial_\mu \varphi 
+ m\, \big(\partial^\nu B_{\nu\mu} \big)\, \phi^\mu, 
\label{92}
\end{eqnarray}
which demonstrate that the kinetic term for the {\it pure} scalar field is positive. It is interesting to point out that {\it both} 
(i.e. the pure scalar and pseudo-scalar) fields obey the normal Klein--Gordan equations of motion, namely; 
\begin{eqnarray}
\big(\Box + m^2 \big)\,\varphi = 0, \qquad \big(\Box + m^2 \big)\, \tilde \varphi = 0.
\label{93}
\end{eqnarray}
Thus, both the fields/particles are endowed with the proper definition of {\it mass}. However, their kinetic terms are 
with opposite signatures.

We would like to point out {\it now} the peculiarities connected with the kinetic terms associated with the vector field $\phi_\mu$ 
and axial-vector field $\tilde \phi_\mu$. First of all, we observe that these kinetic terms are {\it not} invariant under the 
(anti-)BRST symmetry transformations. As a consequence, the field strength tensors 
$\Phi_{\mu\nu} = \partial_\mu \phi_\nu - \partial_\nu \phi_\mu$ and  
$\tilde \Phi_{\mu\nu} = \partial_\mu \tilde \phi_\nu - \partial_\nu \tilde \phi_\mu$ 
are {\it not} gauge-invariant quantities. Thus, these can {\it not} be identified with the $U(1)$ gauge potential $A_\mu$ which is 
present in the field strength tensor $F_{\mu\nu} = \partial_\mu A_\nu - \partial_\nu A_\mu$ of the Maxwell theory (cf. Appendix A below). 
Second point to be noted is the observation that {\it both} of kinetic terms have {\it opposite} signs. Thus, if one of 
them corresponds to an observable field/particle, the other would correspond to  the dark matter because both the 1-form 
potentials obey the proper Klein--Gordon EOM like Eq. (\ref{93}). Hence, both the vector and axial-vector  fields are endowed with 
the proper definition of the {\it rest mass}. However, these real fields have explicit kinetic terms with different 
signatures. Therefore, one of them is one of the possible candidates of dark matter.

We would like to end this section with the concluding remarks that 4D {\it massive} Abelian 2-form gauge theory is a tractable 
field-theoretic model of the Hodge theory which is endowed with multitude of discrete and continuous symmetry 
transformations that provide the physical realizations of {\it all } the mathematical  ingredients associated with
 the de Rham cohomological operators of differential geometry at the {\it algebraic} level. 
In particular, it is the existence of the {\it discrete} symmetry transformations (cf. Eq. (\ref{82})) that provide the physical
 realizations of the Hodge duality operation of differential geometry in $\delta = \pm\, *\, d\,* $. Thus, for the model to be a Hodge theory 
(within the framework of BRST formalism), all the terms of the coupled (but equivalent) Lagrangian densities are {\it fixed}. 
As a consequence, there is {\it no} freedom to change them by any other kind of terms in any manner. Thus, it is the symmetries of the 
field-theoretic model for the Hodge theory that {\it force} the existence of fields/particles with {\it negative} 
kinetic terms which turn out to be the possible  candidates for the dark matter because  their {\it masses} are defined properly. Finally, 
we note that the {\it massless} limit (i.e. $m = 0$), in the St{\"u}ckelberg modified version of Abelian 1-form and 2-form gauge theories, lead 
to the existence of fields/particles with {\it negative} kinetic energy terms {\it only} (cf. our earlier works  \cite{rk1,rpm5}).
Such fields/particles would correspond to the  possible candidates of dark energy within the framework of BRST approach to 
$p$-form ($p = 2, 3, 4,...$) gauge theories (because the basic fields in these theories are taken to be {\it massless} 
due to the power and potential of the gauge symmetries). Hence,
we note that the study of field-theoretic models of Hodge theory are, ultimately,  useful in the physical sense.
 We would like to add here that the fields with negative kinetic term have been christened as the ``ghost" fields in the 
context of self-accelerated, bouncing and cyclic models of the Universe in the realm of cosmology. We make some 
passing comments in the next section on the physical meaning of these fields.


\section{Conclusions} \label{Sect. 9}

In our present investigation, we have shown that the 4D {\it massive}  Abelian 2-form gauge theory is a tractable 
field-theoretic model for the Hodge theory within the framework of BRST formalism (where the celebrated 
St{\"u}ckelberg's approach has been exploited to convert the {\it massive} Abelian 2-form theory into a {\it gauge} theory). 
In the process of the proof of the present model to be an example of Hodge theory, we have been forced to incorporate a pseudo-scalar field and an 
axial-vector field which turn up, in the theory, with {\it negative} kinetic terms but with appropriate definition of {\it mass}. 
Hence, such kind of fields/particles are one of the possible candidates for the dark matter. The massless limit of such fields/particles 
are described by {\it only} the negative kinetic terms. Thus, such {\it massless} fields/particles are one of the possible  candidates for the
dark energy. We, ultimately, conclude that the possible candidates of dark matter and dark energy can be discussed and described in 
a {\it unified} manner within the framework of BRST approach to the $p$-form ($p = 1, 2, 3,...$) {\it massive} theories in 
$\text{D} = 2p$ dimensions of spacetime (where theoretical trick of St{\"u}ckelberg's approach plays an important role).

In the context of the above, it is pertinent to point out that we have proven that the 2D Proca theory, with the help of 
St{\"u}ckelberg's approach, is a model for the Hodge theory within the framework BRST formalism where 
{\it only} a single pseudo-scalar field is incorporated in the theory \cite{rpm10}.  This field turns up with the {\it negative} kinetic 
term {\it but} with a proper definition of mass (because it satisfies the {\it proper} Klein--Gordon equation of motion). 
An essential feature of such kinds of theories is the existence of {\it discrete} symmetry transformations which provide 
the physical realizations of the Hodge duality operation of differential geometry in the relationship: $\delta = \pm\, *\, d \, *$. 
It is the requirement of such kinds 
of symmetries that forces the existence and emergence of fields/particles with {\it negative} kinetic term. The other 
continuous symmetries of the theory provide the physical realizations of the de Rham cohomological operators of differential 
geometry within the framework of BRST approach to {\it massive} $p$-form gauge theories. In fact, the operator form of the bosonic
and fermionic transformations
satisfy the Hodge algebra \cite{eug,mukh,van,des} thereby rendering  the theory to become a model for the Hodge theory.

We would like to lay emphasis on the fact that {\it when} we have considered the 2D free (non-)Abelian 1-form gauge 
theories (without mass) as well as 4D free Abelian 2-form gauge theory (without mass), we have ended up with 
the pseudo-scalar fields with {\it negative} kinetic terms {\it only} (without any mass). Hence, the proof of 
the Abelian $p$-form  ($p = 1, 2, 3, ...$) gauge theories in  $\text{D}= 2p$ dimensions of spacetime (to be a model for the Hodge theory) 
leads to the existence and emergence of the possible candidates of {\it dark} energy (which are characterized by {\it only} 
the negative kinetic terms) \cite{rpm0,rpm5}. However, we have shown  that, in the proof of {\it massive} Abelian $p$-form ($p = 1, 2, 3, ...$) 
theories to be the models for the Hodge theory (within the framework of BRST formalism), the {\it new} fields turn up with the
negative kinetic terms {\it but} with proper definition of {\it mass}. Hence, they are one of the possible candidates of {\it dark matter}.

In the context of various models of accelerated Universe,  the fields with negative kinetic term have been called as the
``ghost" fields which are totally different from the Faddeev-Popov ghost terms of BRST formalism. During the past few years, 
the existence and stability of the vacuum corresponding to the ``ghost" fields have been subject of intense interest in 
the realm of cosmological models \cite{muk1,muk2,ark1,ark2,kro,koy,barr,dva}. These fields have been inevitable in the context of bouncing, self-accelerated 
and cyclic models of the Universe \cite{ste1,ste2,nove,leh,cai1,cai2,alex}. As far as the stability of the vacuum (w.r.t. this field) within the framework 
of our BRST formalism is concerned, there is no problem because the physical state/vacuum is the {\it harmonic} state that is 
annihilated by the BRST and co-BRST charges. Similarly, the unitarity and consistency of our theory is in fine shape 
because of the existence of the off-shell nilpotent and conserved BRST and co-BRST charges. Hence, fields with negative 
kinetic terms do {\it not} create any problem for our physical {\it massive} Abelian 2-form gauge theory and they are 
well-defined physical fields (in our case).

We have proven the free 6D Abelian 3-form gauge theory to be a model for the Hodge theory \cite{rk1}. It would be a nice 
future endeavour to prove the {\it massive} 6D Abelian 3-form gauge theory to be the tractable field-theoretic 
example for the Hodge theory. In this context, we guess that we shall have to incorporate an axial Abelian 2-form field, 
an axial-vector 1-form field and a pseudo-scalar  field (in the St\"{u}kelberg modified version of a {\it massive} Abelian 3-form 
gauge theory) to prove {\it it} to be a model for the Hodge theory. All these {\it new} fields would appear with {\it negative} 
kinetic terms and with proper definition of {\it mass}. As a consequence, all these fields/particles would be the possible candidates 
of dark matter. It is straightforward to draw the conclusion that, in the {\it massless} limit, these fields/particles would 
correspond to the possible candidates for {\it dark} energy. We are actively involved with this problem and our results 
would be reported elsewhere in our future publications.

\section*{Acknowledgements}
RK would like to thank the UGC, Government of India, New Delhi, for financial support under the PDFSS scheme and SK 
would like to gratefully acknowledge the DST 
research grant EMR/2014/000250 for his post-doctoral fellowship. Fruitful and enlightening comments/remarks by our esteemed
Reviewer is thankfully acknowledged. 

\renewcommand{\theequation}{A.\arabic{equation}}    
\setcounter{equation}{0}  


\appendix
\section*{Appendix A: On the discrete symmetries of 2D Proca \\$~~~~~~~~~~~~~~~~~~$ theory: Negative kinetic term}
 \label{Appendix A}

We briefly mention here  the key points connected with the two $(1+1)$-dimensional (2D) Proca (i.e. a $massive$ 2D Abelian 1-form)
theory where the symmetry considerations lead to the  existence and emergence of a pseudo-scalar field with 
{\it negative} kinetic term \cite{rpm10}. In this context, first of all, we begin with the Lagrangian density ${\cal L}_{(0)}^{(P)}$ for the 
Proca theory in any {\it arbitrary} dimension of spacetime (with rest mass $m$) as:
\begin{eqnarray}
{\cal L}_{(0)}^{(P)} = -\frac{1}{4}\, F^{\mu\nu} F_{\mu\nu} + \frac{m^2}{2}\, A^\mu A_\mu,
\label{A1}
\end{eqnarray}
where the 2-form $F^{(2)} = d A^{(1)} \equiv  \frac {1}{2!}\, (dx^{\mu}\wedge dx^{\nu})\, F_{\mu\nu}$ defines the field strength tensor 
$F_{\mu\nu}  = \partial_\mu A_\nu - \partial_\nu A_\mu$ for the {\it Abelian} 1-form $(A^{(1)} = dx^{\mu}A_\mu$) vector gauge field $A_\mu$.
The above Lagrangian density leads to the following EL-EOM (with $m^2 \neq 0$), namely;
\begin{eqnarray}
\partial_\mu F^{\mu\nu} + m^2\, A^\nu = 0 \;\;\Longrightarrow \;\; (\partial \cdot A) \equiv (\partial_\mu A^\mu) =0. 
\label{A2}
\end{eqnarray}
Taking into account the Lorentz gauge $(\partial \cdot A) = 0$, we  observe that we have obtained the Klein--Gordan EOM: 
$(\Box + m^2) \,A_\mu = 0$ for a {\it massive}  Abelian   vector field $A_\mu$. This establishes the fact that the vector 
field $A_\mu$ is a {\it massive} bosonic field. At this stage, there is {\it no} gauge symmetry in the theory as this 
massive Abelian 1-form theory is endowed with the {\it second-class} constraints in the terminology of Dirac's prescription 
for the classification scheme \cite{dira,sund}.  Using the St{\"u}ckelberg approach to {\it massive} gauge theories, 
we modify the Lagrangian density ${\cal L}^{(P)}_{(0)}$ of the Proca theory  with the following re-definitions 
\begin{eqnarray}
A_\mu \to A_\mu \mp \frac{1}{m}\, \partial_\mu \phi, 
\label{A3}
\end{eqnarray}
where $\phi$ is the pure {\it scalar} field. The substitution of this modified form of the vector potential into 
(\ref{A1}) leads to the following modified version of the Lagrangian density
\begin{eqnarray}
{\cal L}_{(0)}^{(P)} \to {\cal L}_{(1)}^{(P)} = -\frac{1}{4}\, F^{\mu\nu} F_{\mu\nu} 
+ \frac{m^2}{2}\, A^\mu A_\mu \mp m \, A_\mu\, \partial^\mu \phi 
+ \frac{1}{2}\, \partial^\mu \phi\, \partial_\mu \phi,
\label{A4} 
\end{eqnarray} 
where the {\it pure} scalar field $\phi$ has the {\it positive} kinetic term. It can be readily checked that (\ref{A4}) 
respects the following gauge symmetry transformations (i.e. $\delta_g  {\cal L}_{(1)}^{(P)} = 0$), namely;
\begin{eqnarray}
\delta_g A_\mu = \partial_\mu \chi, \qquad \delta_g \phi = \pm\, m\, \chi, 
\label{A5}
\end{eqnarray}
where $\chi$ is the local gauge transformation parameter. At this stage, the EL-EOMs, emerging from the Lagrangian density (\ref{A4}),  are 
\begin{eqnarray}
&& \big(\Box + m^2 \big) A_\mu - \partial_\mu (\partial \cdot A) \mp m\, \partial_\mu \phi = 0, \nonumber\\
&& \Box \phi \mp m\, (\partial \cdot A) = 0,
\label{A6}
\end{eqnarray}
w.r.t. the gauge field $A_\mu$ and pure scalar field $\phi$, respectively.  The latter equation can  be $also$ derived from the
former equation by applying an ordinary derivative on it. This form of the Lagrangian density in (\ref{A4}) is 
true for any {\it arbitrary} dimension of spacetime for an Abelian 1-form  vector field $A_\mu$ within the 
framework of St{\"u}ckelberg's formalism.

We now focus on the 2D version of the St{\"u}ckelberg modified Lagrangian density (\ref{A4}) which reduces to the following form
(with $- \frac{1}{4}\, F^{\mu\nu}F_{\mu\nu} = \frac{1}{2}\, E^2$) 
\begin{eqnarray}
{\cal L}^{(P)} &=& \frac {1}{2}\, E^2 + \frac {m^2}{2} A_\mu\, A^ \mu \mp m A_\mu \,\partial^\mu\, \phi 
+ \frac {1}{2}\, \partial_\mu\, \phi\, \partial^\mu\, \phi, 
\label{A7}
 \end{eqnarray}
where $F_{01} = E$ is the electric field for the 2D theory (because this is the $only$ existing competent of the field strength tensor
$F_{\mu\nu}$). It is also clear that $E$ is a pseudo-scalar in two dimensions because it has {\it only} one component and it changes 
sign under parity. This is due to the fact that the electric field $E$ is a polar vector (unlike the magnetic field which is an axial vector).
We note that, in 2D, the mass dimension of $A_\mu$ field is $zero$ (i.e. $[M]^{0}$) as is the case with the
scalar field $\phi$ but the electric field $E$ has the mass dimension equal to {\it  one} (i.e $[M]$) in the natural units: 
$\hbar = c = 1$. For the canonical quantization of our theory (described by the Lagrangian density (\ref{A7}))  as well as for the 
definition of the proper propagator of the {\it ``massive"} gauge field $A_\mu$, we have to incorporate the gauge-fixing term which owes its 
origin to the co-exterior derivative of differential geometry, namely;
\begin{eqnarray}
\delta A^{(1)} = -\;* d * (dx^{\mu}\,A_\mu)  = (\partial\cdot A).
\label{A8}
\end{eqnarray}
It is self-evident that $(\partial\cdot A)$ is a pure scalar and it has the mass dimension of one  (i.e $[M]$).
Hence, we have the freedom to add/subtract a pure scalar field with {\it proper} mass dimension. Such a gauge-fixing term
is: $(\partial\cdot A \pm m\,\phi)$. Thus, the modified Lagrangian density, with the proper gauge-fixing term, is:
\begin{eqnarray}
{\cal L}^{(P)}_{(1)} &=& \frac {1}{2}\, E^2 + \frac {m^2}{2} A_\mu\, A^ \mu \mp m A_\mu \,\partial^\mu\, \phi 
+ \frac {1}{2}\, \partial_\mu\, \phi\, \partial^\mu\, \phi \nonumber\\
&-& \frac {1}{2}(\partial\cdot A \pm m\,\phi)^2. 
\label{A9}
\end{eqnarray}
We now focus on the kinetic term $(-\frac{1}{4}\, F^{\mu\nu} F_{\mu\nu} = \frac {1}{2}\, E^2)$ for the 2D Proca theory.
As pointed out earlier, the field strength tensor $F_{\mu\nu}$ (derived from the 2-form $F^{(2)}  = d\,A^{(1)}$) has 
only $one$ existing component $F_{01} = E$. This field is an  {\it anti-self-dual} field  in 2D. This is due to the
fact that when we apply the  Hodge duality operation   on {\it this} 2-form (with the choice $\varepsilon_{\mu\nu}$ 
as the 2D Levi-Civita tensor and $\varepsilon^{\mu\nu}$ is its inverse), we obtain: 
\begin{eqnarray}
* \,(d A^{(1)}) &=&  * \, \Big [ \frac{1}{2!}\, (dx^{\mu}\wedge dx^{\nu})\, F_{\mu\nu}\Big] \nonumber\\
&=& \frac {1}{2!}\,\varepsilon^{\mu\nu}\,F_{\mu\nu} =  \varepsilon^{\mu\nu} \partial_\mu A_\nu = - E. 
\label{A10}
\end{eqnarray}
Thus, we observe that $(E \to - E)$ under the duality operation in the case of 2D theory. This is a {\it pseudo-scalar}
which can be modified in the following manner (see, e.g. \cite{rpm10})
\begin{eqnarray}
\frac {1} {2}\, E^{2} \to \frac {1}{2}\, (E\mp \,m\,\tilde\phi)^2 
- \frac {1}{2}\,\partial_\mu \tilde \phi\,\partial^{\mu} \tilde \phi \pm m\, E\, \tilde \phi,
\label{A11}
\end{eqnarray}
where  $\tilde\phi$ is a pseudo-scalar field with appropriate kinetic term and an interaction term with the electric field. 
With the above  modification, we have the $final$ form of the Lagrangian density  for the modified version of the 2D Proca 
theory as (see, e.g. \cite{rpm10} for details)     
\begin{eqnarray}
{\cal L}^{(P)}_{(2)} &=& \frac {1}{2}\, {(E \mp m\,\tilde\phi)}^2  -\frac {1}{2}\,\partial_\mu \,\tilde\phi\,\,\partial^\mu\,\tilde\phi 
+\frac {m^2}{2} A_\mu\, A^ \mu \pm m\, E\,\tilde\phi\nonumber\\
&\mp& m\,A_\mu \,\partial^\mu\, \phi + \frac {1}{2}\, \partial_\mu\, \phi\, \partial^\mu\, \phi 
- \frac {1}{2}\,(\partial\cdot A \,\pm\, m\,\phi)^2,
\label{A12}
\end{eqnarray}
which respects the following {\it discrete} symmetry transformations: 
\begin{eqnarray}
A_\mu \rightarrow \pm\,i\,\varepsilon_{\mu\nu}\, A^\nu, \qquad \phi \to \pm\,i\,\tilde \phi, \qquad \tilde \phi \to \mp\,i\, \phi.
\label{A13}
\end{eqnarray}
Thus, we note that, to have the {\it perfect} discrete symmetry in the theory, we have to incorporate a pseudo-scalar field $(\tilde \phi)$
with {\it negative} kinetic term. In fact, the modifications in (\ref{A11}) have been made keeping in mind the discrete symmetry 
transformations (\ref{A13}). We have utilized the discrete symmetry transformation: $A_\mu \to \pm\, i\, \varepsilon_{\mu\nu}\, A^\nu$ 
in our earlier work, too \cite{rpm0,rpm7} where we have discussed the topological nature of 2D (non-)Abelian gauge theories. It is very interesting 
to point out that the mass term $(\frac{m^2}{2}\, A_\mu A^\mu)$ remains invariant under the {\it discrete} symmetry transformations for 
$A_\mu$ as is the case with $(\frac{m^2}{4}\, B_{\mu\nu} B^{\mu\nu})$ for the Abelian 2-form theory under (\ref{22}). It is straightforward to check 
that the {\it pure} scalar and {\it pseudo-scaler} fields obey the Klein--Gordon equation of motion:
\begin{eqnarray}
(\Box + m^2 )\, \phi = 0, \qquad    (\Box + m^2 )\, \tilde \phi = 0. 
\label{A14}
\end{eqnarray}
At this stage, the other field equations are:
\begin{eqnarray}
(\Box + m^2)\, (\partial \cdot A) = 0, \qquad (\Box + m^2)\, E = 0, \qquad
(\Box + m^2)\,A_\mu = 0.  
\label{A15}
\end{eqnarray}
We conclude from (\ref{A14}) that the pseudo-scalar field is a possible candidate for the {\it dark matter} because it possesses the 
{\it negative} kinetic term {\it but} is endowed with the proper definition of  {\it mass} as it satisfies the 
proper Klein--Gordon equation of motion. The {\it discrete} symmetry transformation (\ref{A13}) have been generalized 
(within the framework of BRST formalism applied to the 2D Proca theory) and these symmetries play crucial role 
in providing the physical realizations of the Hodge duality operation of differential geometry \cite{rpm10}.

We end this Appendix with the concluding remarks that the 2D Proca (i.e. a massive Abelian 1-form) theory has
been considered within the framework of BRST formalism and we have shown that the {\it generalized} form of the 
Lagrangian density (\ref{A12}) (that incorporates the Faddeev-Popov ghost terms)  provide a tractable field-theoretic 
model for the Hodge theory where the pseudo-scalar field turns out to be a possible candidate for the dark matter \cite{rpm10}. 
In fact, the continuous and discrete symmetry transformations of the (anti-)BRST invariant Lagrangian densities provide the physical
realizations of the de Rham cohomological operators of differential geometry \cite{eug,mukh,van,des}. In particular, the generalized 
version of the {\it discrete} symmetry transformations (\ref{A13}) provide the physical realizations of the Hodge duality $*$ operation of 
differential geometry (which is {\it one} of the crucial mathematical ingredients of the de Rham cohomological operators because 
the (co-)exterior derivatives are related by $\delta = \pm\, * \,d \,*$). It is the requirement and existence of the {\it discrete} 
symmetry transformations that forces the kinetic term for the pseudo-scalar field to possess a {\it negative} sign and, hence, it 
becomes a possible candidate for the dark matter. The massless limit (i.e. $m = 0$) leads to the existence of {\it dark energy} as, 
in this limit, {\it only} the {\it negative } kinetic term exists.

\renewcommand{\theequation}{B.\arabic{equation}}    
\setcounter{equation}{0}  

\section*{Appendix B: Invariance of the kinetic term}
\label{Appendix B}


We perform here explicit computations connected with the change in the kinetic term $(\frac{1}{12}\, H^{\mu\nu\eta} H_{\mu\nu\eta})$ 
under the modification (cf. Eq. (\ref{2})) where we have the following:
\begin{eqnarray}
B_{\mu\nu} \to B_{\mu\nu} - \frac{1}{m}\, \big(\partial_\mu \phi_\nu - \partial_\nu \phi_\mu 
+ \varepsilon_{\mu\nu\eta\kappa}\, \partial^\eta \tilde \phi^\kappa \big). 
\label{B1}
\end{eqnarray}
It can be explicitly checked that, under  (\ref{B1}), we obtain the following 
\begin{eqnarray}
H_{\mu\nu\eta} \to H_{\mu\nu\eta} - \frac{1}{m}\, \Sigma_{\mu\nu\eta},
\label{B2}
\end{eqnarray}
where the totally antisymmetric tensor $\Sigma_{\mu\nu\eta}$ is explicitly expressed as: 
\begin{eqnarray}
\Sigma_{\mu\nu\eta} = \big(\varepsilon_{\mu\nu\rho\sigma}\, \partial_\eta 
+ \varepsilon_{\nu\eta\rho\sigma}\, \partial_\mu + \varepsilon_{\eta\mu\rho\sigma}\, \partial_\nu \big)\,\partial^\rho \tilde \phi^\sigma.
\label{B3}
\end{eqnarray}
It is now straightforward to check that the kinetic term transforms as: 
\begin{eqnarray}
\frac{1}{12}\, H^{\mu\nu\eta} H_{\mu\nu\eta} \to \frac{1}{12}\, H^{\mu\nu\eta} H_{\mu\nu\eta} 
- \frac{1}{6\,m} H^{\mu\nu\eta} \Sigma_{\mu\nu\eta}  + \frac{1}{12\, m^2}\, \Sigma^{\mu\nu\eta} \Sigma_{\mu\nu\eta}. 
\label{B4}
\end{eqnarray}
At this stage, it is crystal clear that (\ref{B4}) remains trivially invariant under the gauge transformations (\ref{5}) and (\ref{6}). 
As a consequence, the terms on the r.h.s. of (\ref{B4}) remain invariant (due to $s_{(a)b} \,H_{\mu\nu\eta} = 0$ and 
$s_{(a)b}\, \tilde \phi_\mu = 0$) under the (anti-)BRST symmetry transformations (\ref{24}) and (\ref{23}), too. 
It is interesting to state that (\ref{B4}) {\it also} remains invariant under the (anti-)co-BRST symmetry transformations 
(\ref{44}) and (\ref{45}) which lead to the following
\begin{eqnarray}
s_d H_{\mu\nu\eta} &=&  - \big(\varepsilon_{\mu\nu\rho\sigma}  \partial_\eta 
+ \varepsilon_{\nu\eta\rho\sigma}  \partial_\mu 
+ \varepsilon_{\eta\mu\rho\sigma}  \partial_\nu  \big)\big(\partial^\rho \bar C^\sigma\big), \nonumber\\
s_d \Sigma_{\mu\nu\eta} &=&  - m \big(\varepsilon_{\mu\nu\rho\sigma} \partial_\eta 
+ \varepsilon_{\nu\eta\rho\sigma} \partial_\mu 
+ \varepsilon_{\eta\mu\rho\sigma} \partial_\nu  \big)\big(\partial^\rho \bar C^\sigma\big), \nonumber\\
s_{ad} H_{\mu\nu\eta} &=&  - \big(\varepsilon_{\mu\nu\rho\sigma} \partial_\eta 
+ \varepsilon_{\nu\eta\rho\sigma} \partial_\mu 
+ \varepsilon_{\eta\mu\rho\sigma} \partial_\nu  \big)\big(\partial^\rho C^\sigma\big), \nonumber\\
s_{ad} \Sigma_{\mu\nu\eta} &=&  - m \big(\varepsilon_{\mu\nu\rho\sigma} \partial_\eta 
+ \varepsilon_{\nu\eta\rho\sigma} \partial_\mu 
+ \varepsilon_{\eta\mu\rho\sigma} \partial_\nu  \big)\big(\partial^\rho C^\sigma\big). \label{B5}
\end{eqnarray}
We point out, however, that the {\it second} and {\it third} terms in (\ref{B4}) are {\it higher derivative} terms. 
In other words, we note very precisely that there are {\it three} derivatives in the {\it second} term of (\ref{B4}) and there are {\it four} 
derivatives in the {\it third} term. Such terms are problematic and pathological as far as the renormalizability of our present  
theory is concerned. Thus, we do {\it not} consider them in our cohomological discussions. As far as the proof of our present
{\it massive} field-theoretic model, to be an example of the Hodge  theory, is concerned, we focus only on the
Lagrangian densities (\ref{28}) and (\ref{29}) within the framework of BRST formalism which respect {\it discrete} as well as 
{\it continuous} symmetry transformations of various kinds which enable us to  figure and find out the physical realizations 
of the de Rham cohomological operators of differential geometry \cite{eug,mukh,van,des} in terms of the conserved charges.


\section*{Appendix C: Emergence of CF-type restrictions}
 \label{Appendix C}
The existence of the CF-type restriction(s) is the hallmark of a $p$-form gauge theory when it is discussed within the framework 
of BRST formalism \cite{bon1,bon2}. To be more precise, at the {\it quantum} level, the existence of the CF-type restriction(s) is as 
fundamental as the existence of the first-class constraints at the {\it classical} level for a given $p$-form gauge theory. 
In our present Appendix, we show the emergence of CF-type restrictions using a diagram where a {\it single} field is denoted 
by a {\it single} circle at a point and the (anti-)BRST symmetry transformations $s_{(a)b}$ have been shown by arrows corresponding 
to the transformations listed in Eqs. (\ref{24}) and (\ref{23}) where a specific field transforms to another field.

In the diagram, there are {\it two} layers of fields. The fields at the top layer are represented by the blue circles and they 
correspond to all the fields that are obtained after the application of (anti-)BRST symmetry transformations on the Abelian 
2-form $B_{\mu\nu}$ field (and its descendants). The blue arrows denote the (anti-)BRST symmetry transformation operators 
$s_{(a)b}$. There is a bottom layer which corresponds to the Abelian 1-form field $\phi_\mu$ and its descendants that are obtained 
by the application of (anti-)BRST symmetry transformations $s_{(a)b}$. The bottom fields {\it also} transform to 
{\it top layer} fields by the (anti-)BRST symmetry operators $s_{(a)b}$. The latter are denoted by the red arrows. The operation of exterior 
derivative $d$ lifts the lower ranked fields to higher rank fields. In other words, we note that $s_b B^{(2)} = d C^{(1)}$ 
and $s_{ab} B^{(2)} = d \bar C^{(1)}$ which imply $s_b B_{\mu\nu} = - (\partial_\mu C_\nu - \partial_\nu C_\mu)$ and
 $s_{ab} B_{\mu\nu} = - (\partial_\mu \bar C_\nu - \partial_\nu \bar C_\mu)$ modulo a sign factor. The key CF-type restriction: 
$B_\mu + \bar B_\mu + \partial_\mu \varphi = 0$ is also connected by the relationship: $B^{(1)} + \bar B^{(1)} + d \varphi^{(0)} = 0$
where the 0-form scalar field $\varphi$ is lifted to the 1-form fields ($B^{(1)} = dx^\mu B_\mu$ and   
$\bar B^{(1)} = dx^\mu \bar B_\mu$) by the application of exterior derivative $d$.

\begin{figure*}[ht]
\begin{center}
\includegraphics[scale=.68]{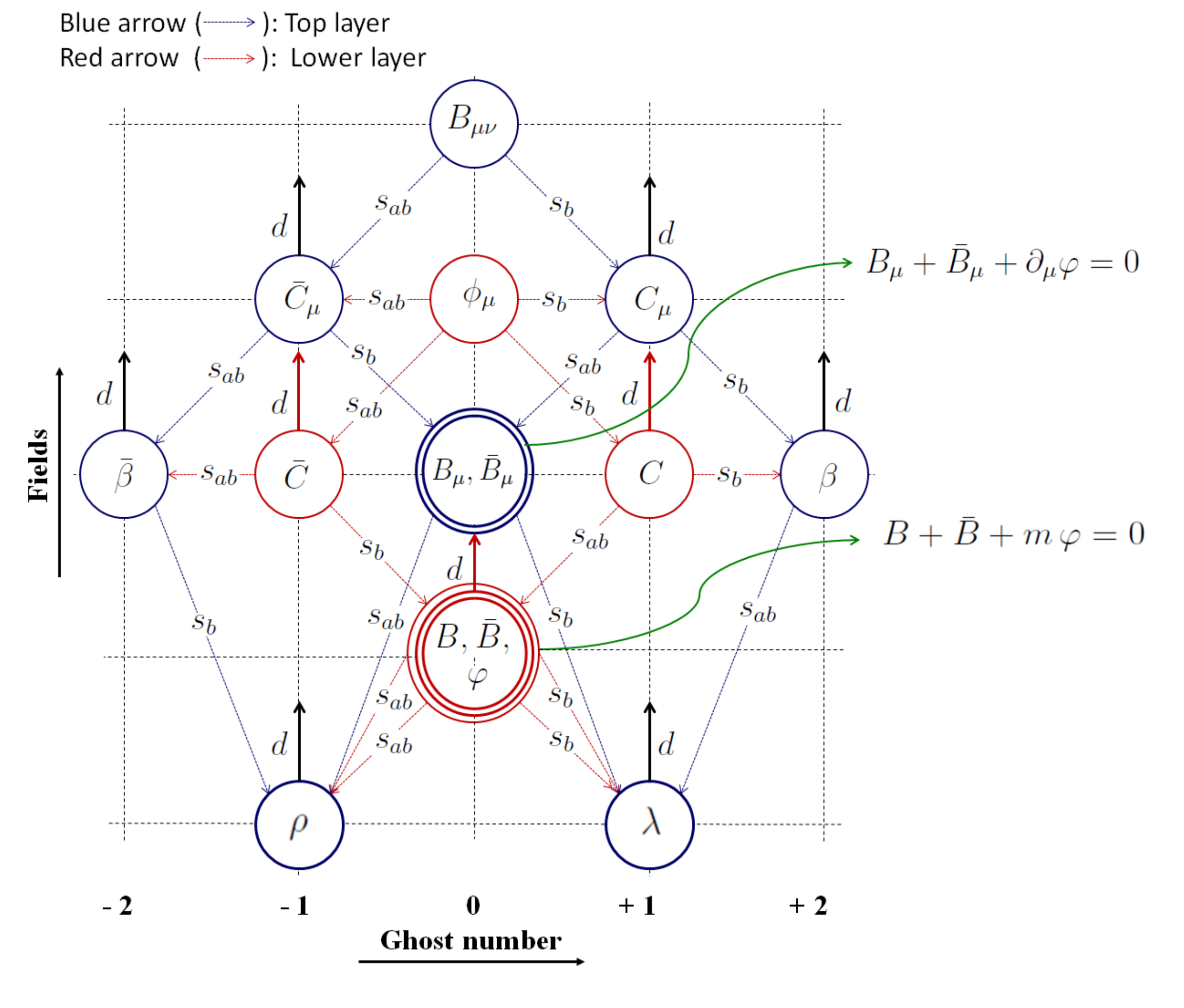}
\caption{Emergence of CF-type conditions.} \label{Fig.1}
\end{center} 
\end{figure*}

\begin{figure*}[ht]
\begin{center}
\includegraphics[scale=0.68]{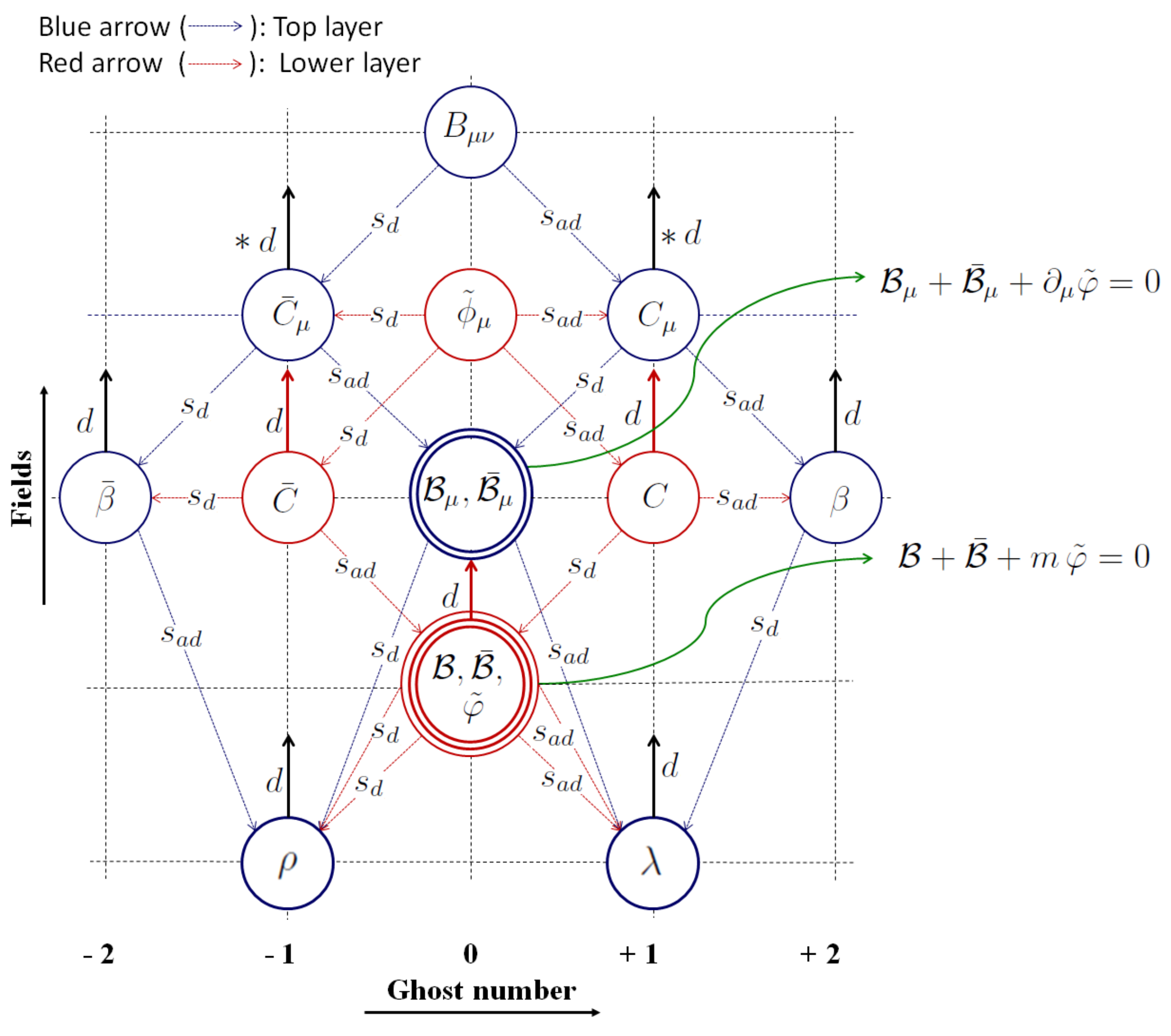} 
\caption{Emergence of CF-type conditions.} \label{Fig.2}
\end{center} 
\end{figure*}

The key observation of diagram (cf. Fig.\ref{Fig.1}) is the fact that whenever {\it two} or {\it three} fields cluster at the same point, 
there would be the existence  of Cf-type restrictions where (i) either {\it three} fields, existing in the same plane, would 
be connected by a restriction (i.e. $B + \bar B + m \varphi = 0$), (ii) or {\it two} fields in the same plane (i.e. $B_\mu$ and 
$\bar B_\mu$) would be connected to a lower rank field (i.e. $\varphi$ existing in the bottom plane) by an exterior derivative 
(i.e. $B_\mu + \bar B_\mu + \partial_\mu \varphi = 0$). The clustering of the fields has been denoted by double concentric 
circles and/or {\it triple} concentric circles in the diagram.         
In the above paragraph, we have discussed  the possible existence of CF-type restrictions in the case of 
(anti-)BRST symmetry transformations through the diagram (cf. Fig. \ref{Fig.1}) and demonstrated that the clustering 
of fields at a point (with the same ghost number) ensures the emergence of CF-type restrictions/conditions (which 
are the decisive features of a {\it quantum} gauge theory within the framework of BRST formalism). In our present theory, 
there are nilpotent and absolutely anticommuting (anti-)co-BRST symmetry transformations, too.  These latter symmetries are 
{\it absolutely} anticommuting {\it only} on a hypersurface in the 4D Minkowaskian spacetime manifold where the CF-type 
restrictions (i.e. $ {\cal B}_\mu + \bar {\cal B}_\mu + \partial_\mu \tilde \varphi = 0$, ${\cal B} + \bar {\cal B} + m \tilde \varphi = 0$) 
are satisfied. Diagrammatically (cf. Fig. \ref{Fig.2}), the emergence of such kind of restrictions can be {\it also} discussed along exactly similar 
lines of arguments  as we have demonstrated, the emergence of the CF-type restrictions, in the context of 
(anti-)BRST symmetry transformations (Fig. \ref{Fig.1}). There is a decisive and distinct difference, however. We note that, 
in the case of (anti-)co-BRST symmetry transformations (i.e. $s_{(a)d} B_{\mu\nu}$) for the Abelian 2-form field, we have 
the relationships: $s_d B^{(2)} = * d \bar C^{(1)}$ and $s_{ad} B^{(2)} = * d C^{(1)}$ 
(where $B^{(2)} = \frac{1}{2!} (dx^\mu \wedge dx^\nu) B_{\mu\nu}$, $C^{(1)} = dx^\mu C_\mu$ and 
$\bar C^{(1)} = dx^\mu \bar C_\mu$) which are different from the (anti-)BRST symmetry transformations 
where we have: $s_b B_{\mu\nu} = d C^{(1)}$ and $s_{ab} B_{\mu\nu} = d \bar C^{(1)}$.      
In Our earlier works \cite{bon1,bon2}, we have established the connection between the CF-type restrictions and the 
geometrical objects called gerbes. This deep connections {\it physically} imply the linear independence of the BRST 
and anti-BRST symmetry transformations and their corresponding BRST and anti-BRST charges. A similar kind of mathematical 
connection can be established for the CF-type restrictions, existing in the case of (anti-)co-BRST symmetry transformations 
(and their corresponding charges) and the ideas of gerbes. We are working in this direction and our results would be 
reported elsewhere \cite{rk5}.






\end{document}